\documentclass[twocolumn,aps,pra,showpacs]{revtex4}

\usepackage{epsfig}
\newcommand{\bfr}{{\bf r}}

\newcommand{\bfk}{{\bf k}}

\newcommand{\hpsi}{\hat{\psi}}

\newcommand{\hPsi}{\hat{\Psi}}
\newcommand{\ddt}{\frac{d}{dt}}

\begin{document}


\title{Creation, detection and decoherence of Schr\"{o}dinger cat states in Bose-Einstein condensates}



\author{Y. P.  Huang}
\author{M. G. Moore}
\affiliation{Department of Physics \& Astronomy, Ohio University, Athens, OH  45701}


\date{\today}

\begin{abstract}
We study the possibility to create many-particle Schr\"odinger cat-like states by using a Feshbach resonance to reverse the sign of the scattering length of a Bose-Einstein condensate trapped in a double-well potential. To address the issue of experimental verification of coherence in the cat-like state, we study the revival of the initial condensate state in the presence of environmentally-induced decoherence. As a source of decoherence, we consider the interaction between the atoms and the electromagnetic vacuum, due to the polarization induced by an incident laser field. We find that the resulting decoherence is directly related to the rate at which spontaneously scattered photons carry away sufficient information to distinguish between the two atom-distributions which make-up the cat state. We show that for a 'perfect' cat-state, a single scattered photon will bring about a collapse of the superposition, while a less-than-perfect cat-like state can survive multiple scatterings before collapse occurs. In addition, we study the dephasing effect of atom-atom collisions on the cat-like states.\end{abstract}


\maketitle

\section{introduction}
Quantum degenerate gases of bosonic and/or fermionic atoms have proven very useful tools in the exploration of low-temperature many-body quantum physics. In this area of research,  the initial Bose-Einstein condensate (BEC) or degenerate Fermi gas (DFG) serves primarily as a well-defined starting point for the generation of more exotic highly-correlated and/or entangled many-body states.  The most profound recent experimental demonstrations of this approach include the observations of the superfluid to Mott-insulator transition in an optical lattice and the cross-over to a `fermionized' Tonks-Girardeau under quasi-one-dimensional confinement.  In addition, there has been significant progress towards observing the BEC to BCS cross-over in a two-component gas of Fermionic atoms \cite{jin,Holland}, and efforts are underway to observe the cross-over from a vortex-lattice to a non-trivial entangled many-body state in the ground-state of a rapidly rotating BEC \cite{jasonho,ketterle}. 

In the first of the aforementioned experiments, a reversible change from a superfluid to a Mott-insulator state was observed when crossing from  tunneling-dominated to the collision-dominated regime for atoms trapped in a lattice potential with repulsive atom-atom interactions. The analogous transition for the case of attractive interactions involves the crossover from a superfluid into a state often described as a `Schr\"odinger cat state' where the atomic population collapses into a single lattice site, with the true ground state being a symmetric superposition over all possible lattice sites as the final occupied site. In addition to providing fundamental insights into the nature of the transition from the quantum to the classical descriptions of reality, such states may have important applications in precision measurement \cite{wineland} and quantum information processing. 

In this paper we investigate theoretically the possibility to use a BEC trapped in a double-well potiential to generate highly-entangled Schr\"odinger cat-type state, which can be viewed as a realization of the two-site Bose-Hubbard model with attractive interactions. We address important issues such as detection of cat-like states and the effects of photon-scattering induced decoherence which occurs  when far-off-resonant laser-light is used for trapping and/or probing the cat states. 

Interest in Schr\"odinger cat-like states, loosely defined as quantum superpositions of macroscopically distinguishable many-body states, goes back to the earliest days of quantum mechanics. In 1935, in oder to demonstrate the limitations inherent in using quantum mechanics to describe everyday phenomena, Erwin Schrodinger proposed a experiment in the macroscopic state of a cat is entangled with that of an unstable nuclei, allowing the cat to enter a coherent superposition of being dead and alive \cite{Schrodinger}. Such states were thought at the time to be logically untenable, although today we might consider only that they would not be observed due to rapid environmentally-induced decoherence \cite{Zurek1991}. To this day, this  famous gedanken experiment illustrates the fundamental difficulties inherent postulating a well-defined boundary between a 'classical' level of reality and an underlying level governed by quantum mechanics, one of the fundamental open questions in quantum theory.

Many of the early gedanken experiments, whose consideration led to the earliest understanding of quantum mechanics, are now being transformed into laboratory realities. One such effort has been the ongoing quest for Schr\"odinger-cat-like states, with the most striking experimental results to date being the observation of a superposition between clock-wise and counter-clock-wise currents in a superconducting quantum interference device \cite{Jonathan2000}  and the observation of double-slit diffraction for massive $C_{60}$ molecules \cite{Arndt2001}. Additional efforts include work in  trapped ions at low temperature \cite{Monroe1996,Sackett2000}, nanoscale magnets\cite{Friedman1996}, and superconductors \cite{Rouse1995,Silvestrini1997,Nakamura1999}. This progress is generating deeper insight into the meaning of quantum theory, as well as potential applications in the fields of precision measurement and quantum information processing. Even with the current rapid rate of progress, the quest to obtain and utilize large cat-like states remains a serious experimental challenge, due primarily to the presumed scaling of fragility (with respect to environmentally-induced decoherence) with number of particles. For most appliciations, the utility of the cat-state increases with particle number \cite{Osborne,Bose}, so there is a strong need to find systems in which such maximally-entangled states of large numbers of particles can be generated without rapid collapse due to decoherence. 

An atomic BEC \cite{Anderson1995,Bradley1995,Davis1995} may be such a system, as its large degree of spatial and temporal coherence is well established by experiment \cite{dalfovo1999}.  The bimodal BEC, which is constructed by isolating a pair of weakly-coupled atomic-field modes, is one of the simplest quantum systems used in the study of entanglement physics and has therefore been the subject of extensive theoretical study \cite{Leggett2001, Andrew2003}. In this system, $N$ bosons are restricted to occupy one of two modes, and entanglement is established via 2-body interactions between the bosons. For the case of an atomic BEC, the two modes can either be two potential-wells spatially separated by a potential barrier (so-called Double-well BEC's) \cite{Raghavan2, Mahmud2003} or two spatially-overlapping modes in different hyperfine states \cite{Micheli2003,Sorenson2001}. In addition, a recent paper proposes an analogous system in which a double-well `potential' in momentum-space, formed from the band-structure in a lattice potential, can be used to create cat-states involving two momentum-states  \cite{stamperkurn}. 

The quantum ground-state of the bosonic double-well system has been studied for the case of repulsive interactions in the context of the superfluid to Mott-insulator transition \cite{Greiner2002}, as well as for the case of attractive interactions. For the latter, it has been shown under certain conditions a Sch\"{o}dinger cat-like macroscopic superposition state, can be generated\cite{Cirac1998, Ho2000}. In this cat-like many-body state, a large number of atoms are collectively either in one mode or in another.  The dynamics of the bimodal system have also been investigated, both via a meanfield approach and a full many-body treatment within the two-mode approximation. In the mean-field treatment,  the dynamic evolution is treated as a Bose Josephson Junction, which is complementary to that of a superconductor tunnel junction \cite{Smerzi1997, Raghavan1}. This treatment predicts a degenerate ground-state, corresponding to all amplitude in one well or the other, but is not capable of predicting the dynamical evolution of cat-like states, as these states lie outside of the domain of mean-field theory.  In the two-mode many-body treatment, the many-particle cat-like state is predicted to arise during the dynamical evolution of an initial minimally-entangled condensate state in the presence of repulsive interactions \cite{Micheli2003,Mahmud2003}, but with the sign of the tunneling coefficient reversed by imprinting a $\pi$-phase shift onto one well. In this paper, we propose that such cat states can be realized for the case of an atomic BEC trapped in a double-well potential by employing a magnetic Feshbach resonance to tune the atomic interaction from repulsive to attractive \cite{FeshbachWiley1992,Kokkelmans2002}.  It is found that for suitable parameters, the initial state will evolve into a cat-like state after a certain period of time, and then it will revive back to the initial state.

Despite these theoretical advances, there remain many challenges, some of which we address in the present manuscript. One challenge is the problem of experimental verification of the coherence between macroscopically distinguishable states, which is related to the challenge of understanding the role played by environmentally-induced decoherence in collapsing the cat-state onto a statistical mixture of the two macroscopically-distinguishable possibilities. In this paper we focus in detail on the decoherence which will occur if laser-light is used for trapping and/or probing of the cat states. In addition the influence of atom-atom interactions on the proposed detection scheme is studied in some detail.  As suggested in \cite{Diego2000}, we show the revival of the initial state can be used as an unambiguous signal that the supposed cat-state is indeed a coherent macroscopic superposition, as opposed to an incoherent mixture. We also demonstrate, however, that collisional de-phasing can mask the revival without a true loss of coherence.

The effects of decoherence on these cat-like states have primarily been discussed for the case of coupling to the thermal-cloud of non-condensate atoms. There is disagreement in the literature regarding whether or not this effect is significant \cite{Diego2000,Louis,Micheli2003}. We note, however, that these authors treat the non-condensate fraction as a Markovian reservoir, which implies an infinitely-short memory (relative to condensate evolution times)  for correlations between the condensate and non-condensate terms. We question this assumption, primarily due to the fact that the non-condensate fraction typically has a very low temperature and density. In addition, the non-condensate atoms are trapped inside the condensate volume by the trapping potential, thus there is no irreversible loss of information due to propagation of the entangled atoms outside of the system volume. Hence, we suspect that a non-Markovian treatment is necessary to accurately evaluate the thermal-cloud induced decoherence. However, modeling this is a difficult task that we do not address at present. Instead, we assume that this effect is negligible and concentrate primarily on an analysis of a new source of decoherence: that due to spontaneous scattering of far-off-resonant photons from laser fields which we presume are used in either the trapping and/or probing of the system. In this case, the correlation time of the reservoir  is governed by the irreversible loss of the scattered photons from the system volume at the extremely fast rate of $L/c$ where $L$ is the atomic mode size and $c$ is the speed of light. Hence, the Markov approximation works extremely well for describing the physics of photon-scattering. As is well known, this approach leads to effective interactions as well as decoherence, effects which include induced dipole-dipole interactions and collective spontaneous emission (superradiance) in the atomic ensemble.

When analyzing the effects of environmentally-induced decoherence we are interested in answering two basic questions: (1) What  limitations are imposed on the ability to create cat-states, and revive the initial condensate states by the presence of decoherence, and (2) how long can one `hold' the cat-state  by decreasing the tunnel-coupling to zero, without collapse of the cat-state onto a mixed state. The second question is raised in anticipation of applications in precision measurement and quantum information processing where the cat state is operated on and/or held ready for future operations during a complicated protocol. We therefore need to determine the lifetime of the cat states in the presence of decoherence, as well as the effect of decoherence on the dynamical evolution from the initial state into the cat-state. 

The organization of this article is as follows.  In Section \ref{model}, the model of bimodal BEC in Double-well system is introduced and the cat states in such a system are defined; in Section \ref{generation}, two schemes of creating cat state are discussed: adiabatic evolution of the ground-state and and dynamical evolution of a non-stationary state. We demonstrate that the former is experimental unfeasible due to the fact that the end state is nearly degenerate with the first-excited state. We then propose generating a cat-like state through dynamic evolution following a sudden flipping of the sign of the atomic interaction, accomplished via  Feshbach resonance.  In Section \ref{detection}, the method of detecting the cat-state via revival of the initial state is discussed in detail. In Section \ref{decoherence} we investigate the decoherence  of cat-states and a master equation is derived for laser-induced decoherence. In addition, the lifetime of cat-states in the presence of laser fields is determined. In Section \ref{dephasing}, the dephasing effects due to the nonlinear two-body interaction is studied. And finally, we briefly summarize our results in Section \ref{conclusion}.

\section{The model}
\label{model}
Our proposed scheme involves loading a BEC into a double-well potential and then employing a Feschbach resonance to vary the atomic scattering length  in order to produce a Schr\"odinger cat-like state. For present we consider only the case of a spatial double-well potential.  However, an analogous system can be formed by using a Raman-scheme to couple two hyperfine states in a spinor BEC. The primary difference between this model and the double-well system lies in the presence of collisional interactions between the two modes, due to the spatial overlap of the hyperfine modes. Our current model neglects these collisions, as they are negligible for the case of well-separated potential minima. Most of our results should apply to both systems, however, the effects of inter-mode collisions will be considered in future work. 

We begin our analysis from the usual many-body Hamitonian describing atomic BEC,
\begin{eqnarray}
\label{many-body}
	\hat{H}&=&\int d^3r
	\hat{\Psi}^{\dagger}(\bfr)[-\frac{\hbar^2}{2m}\nabla^2+V_{trap}(\bfr)]\hat{\Psi}(\bfr)\nonumber\\
	&+&\frac{U_0}{2}\int d^3r\, \hat{\Psi}^{\dagger}(\bfr)\hat{\Psi}^{\dagger}(\bfr)\hat{\Psi}(\bfr)\hat{\Psi}(\bfr),
\end{eqnarray}
where $m$ is the atomic mass, $V_{trap}$ is the trapping potential and $\hat{\Psi}$ is the annihilate field operator for atoms in
Heisenberg picture. The two-body interaction strength is given by $U_0=4\pi a\hbar^2/m$, with $a$ denoting the $s-$wave scattering
length. In our model, the BEC is assumed to occupy a symmetric double-well potential. 

The single-atom ground state and first excited state this system we denote as $\psi_s(\bfr)$ and $\psi_a(\bfr)$, respectively, where $\psi_s(\bfr)$ is symmetric with respect to the double-well symmetry and $\psi_a(\bfr)$ is antisymmetric. We denote the energy gap between these states as $\hbar\tau$, which is typically much smaller than the gap between the first and second excited states. Because of this gap, we assume throughout that the atomic population is restricted to these two modes alone.
It is convenient to introduce two localized states $\psi_L$, $\psi_R$ \cite{Milburn1997},
\begin{eqnarray}
	\psi_L(\bfr)&=&\frac{1}{\sqrt{2}}\left[\psi_s(\bfr)+\psi_a(\bfr)\right], \nonumber \\
	\psi_R(\bfr)&=&\frac{1}{\sqrt{2}}\left[\psi_s(\bfr)-\psi_a(\bfr)\right].
\end{eqnarray}
Expanding the field operator $\hat{\Psi}(\bfr)$ onto these two modes gives
\begin{equation}
\label{Psiexpand}
	\hat{\Psi}(\bfr,t)=\hat{c}_L(t)\psi_L(\bfr)+\hat{c}_R(t)\psi_R(\bfr),
\end{equation}	
where $\hat{c}_L$ and $\hat{c}_R$ are bosonic atom-annihilation operators for the left and right modes, respectively.
Inserting this expansion back into the many-body Hamiltonian (\ref{many-body}), one recovers the two-mode version of Bose-Hubbard model
\begin{equation}
\label{Bose-Hubbard}
	\hat{H}=-\hbar\tau(\hat{c}^\dagger_L\hat{c}_R+\hat{c}^\dagger_R\hat{c}_L)+\hbar 
	g(\hat{c}^{\dagger 2}_L\hat{c}^2_L+\hat{c}^{\dagger 2}_R\hat{c}^2_R).
\end{equation}
In deriving this expression, we have implicitly assumed that atomic collisions in the overlapping region of the two modes are
negilibile, valid under the assumption that the two modes are well separated.  The intra-well two-body interaction strength is indicated by $ g$, where $g=U_0/2\hbar \int d^3r|\psi_L(\bfr)|^4$. Note that in the deriving of the bimodal Hamiltonian (\ref{Bose-Hubbard}), the wavefunction for each mode is assumed to be independent of particle number in the trap, and terms proportional to the total atom number-operator have been dropped, as atom number is assumed to be a conserved quantity.

The quantum state of this bimodal system can be expessed as a superposition of Fock-states $|n\rangle$ \cite{Andrew2003}
\begin{equation}
\label{eqn-Fock}
	|\hat{\Psi}\rangle=\sum^{N/2}_{n=-N/2}c_n|n\rangle,
\end{equation}
where
\begin{equation}
\label{n}
	|n\rangle=\frac{(c^{\dagger}_L)^{N/2+n}(c^{\dagger}_R)^{N/2-n}}{\sqrt{(N/2+n)!(N/2-n)!}}|\texttt{0}\rangle.
\end{equation}
Here, $N$ is the total number of atoms in condensate which for convenience we take as a even number, and $n$ denotes half number difference between the two modes. The two corresponding operators are
\begin{eqnarray}
\label{Nn}
	\hat{N}=\hat{c}^\dagger_L\hat{c}_L+\hat{c}^\dagger_R\hat{c}_R, \nonumber\\
	\hat{n}=\frac{1}{2}(\hat{c}^\dagger_L\hat{c}_L-\hat{c}^\dagger_R\hat{c}_R).
\end{eqnarray}
In this representation, the conventional Schr\"odinger cat state is defined as
\begin{equation}
\label {cat-define}
	|\texttt{cat}\rangle=\frac{1}{\sqrt{2}}(|N/2\rangle+|-N/2\rangle),
\end{equation}
where the $N$ particles have equal probability to be all in the left or all in the right mode. This cat-state is an ideal maximally-entangled state, however it will be difficult to create such a state in this bimodal system. Instead, a more experimental accessible cat-like state is one containing two-well-separated wave-packets in the distribution of $c_n$. For example, a typical cat-like state can be of the form
\begin{equation}
\label{cat-Gaussian}
	c_n\propto e^{\frac{(n-n_0)^2}{2\sigma^2}}+e^{\frac{(n+n_0)^2}{2\sigma^2}},
\end{equation}
with $n_0>\sigma$.

In the present bimodal system the class of cat-like states is more general than the Gaussian form
(\ref{cat-Gaussian}). Instead, a useful cat like state only need to satisfy two conditions: (i) the probability distribution in Fock-space
should be approximately symmetric; and (ii) the two wave-packets should be well separated in order to correspond to macroscopically distinguishable states. In the present model, condition (i) is satisfied provided the initial state is symmetric, due to the symmetry of Hamiltonian (\ref{Bose-Hubbard}). As we will see is Section \ref{decoherence}, decoherence may break this symmetry. Nonetheless, to
distinguish cat-like states from non-cat state, it is convenient to introduce a projection operator $\hat{P}_{cat}$
\begin{equation}
\label{projection-cat}
	\hat{P}_{cat}=\sum_{n=n_c}^{N/2}(|n\rangle\langle n|+|-n\rangle\langle-n|).
\end{equation}
This operator acts as a filter which picks up cat-states with a minimum wave-packet separation of $2n_c$. And the expectation value $\langle\hat{P}_{cat}\rangle$ can be a efficient way to determine whether a state is cat-like. It is straightforward to see that for a cat-like state we will have $\langle \hat{P}_{cat}\rangle\approx 1$, whereas for a non-cat state the trace will be noticeably less than unity.

\section{Generation of cat states}
\label{generation}

As shown both theoretically and experimentally , the $s$-wave scattering length of cold atoms can be modulated by applying a varied magnetic field. In the  vicinity of a Feshbach resonance, it takes the form
\begin{equation}
\label{Feshbach}
	a=a_{0}(1-\frac{\Delta B}{B-B_{0}}),
\end{equation}
where $a_{0}$ is the background value, $B_{0}$ is the resonant value of the magnetic field, and $\Delta B$ is the width of the resonance. By tuning the magnetic field $B$, the scattering length $a$ can be efficiently tuned in an adequate range based on the Feshbach resonance, as demonstrated in recent experiments. Thus, the two body interaction term
$g$ in Hamiltonian (\ref{Bose-Hubbard}), which is determined by scattering length $a$, is implicitly assumed to be an arbitrary adjustable number.

Before discussing the methods of generating cat-states, we note that attractive BEC's can collapse due to instability, as been demonstrated in experiments \cite{Chin2003}. A trapped BEC, however, can be stable against collapse under the condition of low atomic density \cite{Kagan1997}. In terms of the BEC atom number $N$, the attractive interaction strength $g$ and the energy gap between the first-excited state and ground state $\Delta\omega$,  the stability condition can be written as
\begin{equation}
N<\frac{\Delta\omega}{|g|}.
\end{equation}
In the present discussion one can increase the energy gap between the ground and excited state by adjusting the trapping potential, and in this way, we can ensure our system will not collapse when the sign of the scattering length is switched from positive to negative. 

\subsection{Adiabatic Evolution} 
In this subsection, we begin by briefly discussing the possibility to create cat-states via adiabatic manipulation of the many-body ground state \cite{Cirac1998, Ho2000}, achieved by a continuous variation of the interaction strength $g$ from a positive to a negative value. 
In order to better understand the adiabatic evolution of the ground state, we first examine the ground state under certain parameter values for which it can be determined exactly.

The first case we consider is that of zero tunneling, in which case the bimodal Hamiltonian (\ref{Bose-Hubbard}) can be reduced to
\begin{equation}
\label{no-tunneling}
	\hat{H}=2\hbar g\, \hat{n}^{2}.
\end{equation}
The ground state of this Hamiltonian is determined solely by the parameter $ g$. For
$ g>0$, corresponding to repulsive interactions, the ground state will be $|0\rangle$ in our number-difference representation (\ref{n}),
i.e., an insulator state. While in the case of $ g<0$, the ground state is doubly degenerate, corresponding to any arbitrary superposition of the two extreme states
$|N/2\rangle$ and $|-N/2\rangle$, and its orthogonal counterpart. These states can be written as 
\begin{equation}
	\frac{1}{\sqrt{\alpha^2+\beta^2}}(\alpha|N/2\rangle+\beta|-N/2\rangle,
\end{equation}
 with $\alpha, \beta$ being arbitrary complex numbers. Note when $\alpha=\beta$, one recovers the Shr\"{o}dinger cat state (\ref{cat-define}).

In the other extreme case, if there is no atomic interaction but only  the tunneling term, the ground state of Hamiltonian (\ref{Bose-Hubbard}) becomes the two-site analogue of the superfluid phase \cite{Andrew2003}
\begin{equation}
	|g\rangle=2^{-N/2}\sum^{N/2}_{-N/2}\sqrt{\left(\begin{array}{c}N \\n\end{array}\right)}|n\rangle,
\end{equation}
which is a binomial distribution of Fock states, and thus approximately Gaussian when $N$ is a large number. Henceforth we shall refer to this class of states as a `coherent state', meaning that each atom is in a coherent superposition of both modes and there are no atom-atom correlations or many-body entanglement.

The more general ground state of our model, with the presence of both collisions and tunneling, can be understood as a smooth cross-over between these two extremes. First, under the condition of a repulsive atomic interaction, the ground states of this system can be well approximated by a single Gaussian-like distribution of $c_n$  peaked at $n=0$
\begin{equation}
\label{eqn-Guass}
	c_n\propto e^{-n^2/(2\sigma^2)},
\end{equation}
where the relative atom number dispersion $\sigma$ is obtained by minimizing by the mean-energy of Hamiltonian. Second, under the condition of attractive two-body interaction, the ground state of this system will be split into two separated wave-packets. Thus instead of (\ref{eqn-Guass}), a good approximation of this state turns out to be a superposition of two Gaussian distributions (\ref{cat-Gaussian}) \cite{Ho2000}, where again the minima center $n_0$ and spread width $\sigma$ are obtained by minimizing the mean-energy. For repulsive atomic interactions, the energy minima is found at $n_0=0$, which indicates that the atom number distribution does not split. For attractive interactions, $n_0\neq 0$, and under the condition of $n_0>\sigma$, the wavepacket of ground state will split and become a superposition of two well-separated components, i.e. it becomes a cat-like state (\ref{cat-Gaussian}). Note for this state, the relative phase for any Fock basis pair $|n\rangle$ and $|-n\rangle$ is zero.

Based on these considerations, it is of theoretical interest to determine whether or not  the Shr\"{o}dinger cat state in this bimodal BEC system can be generated by an adiabatic process \cite{Ho2000, Cirac1998}. As a basic scheme, the system would be first  prepared in the many-body ground state with positive $g$, corresponding to a coherent state centered at $n=0$. Then by slowly tuning $g$ to be negative, the distribution will split and finally end up with a Shr\"{o}dinger cat like state under the condition of $n_0>\sigma$. 

\begin{figure}
\epsfig{figure=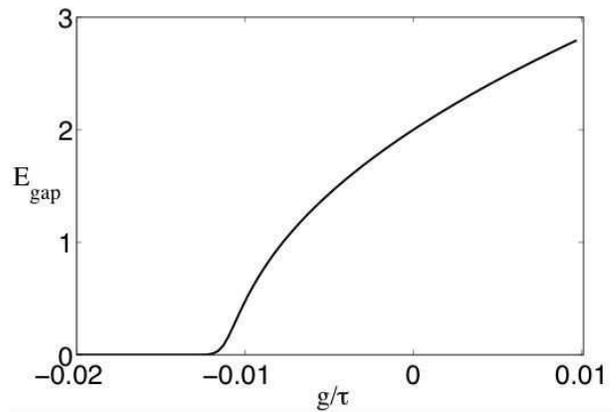, width=8.0cm}
\caption{The energy gap $E_{gap}$ as a function of ratio $g/\tau$. Here, $E_{gap}=E_1-E_0$, with $E_0, ~E_1$ being the eigen energies of the ground and first excited state respectively. \label{degenerate}}.
\end{figure}

\begin{figure}
\epsfig{figure=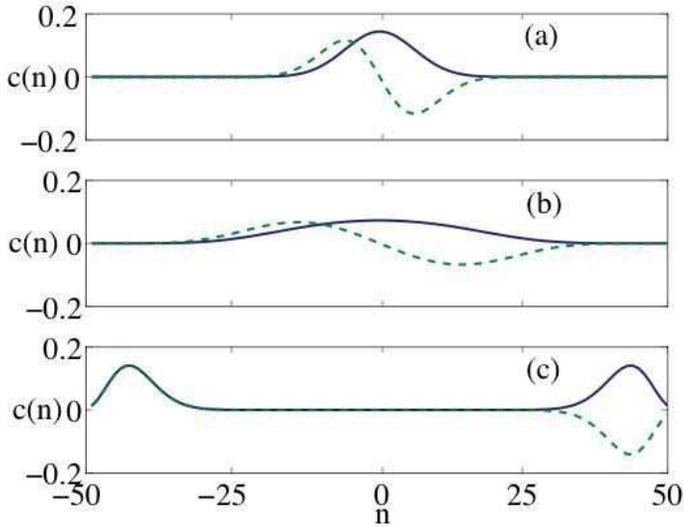, width=9.0cm}
\caption{The wave-packet $c(n)$ of the ground state (Blue solid) and first excited sate (Green dashed) with (a) $g/\tau=0.01$, (b) $g/\tau=-0.01$ and (c) $g/\tau=-0.02$ respectively. \label{degwf}}.
\end{figure}

There is a significant difficulty with this proposal, however, in that when the two wavepackets in Fock space are well separated, the ground state, which is the desired cat state, is nearly degenerate with the first excited state, as shown in Fig.(\ref{degenerate}). It is observed that when $g$ is positive, there is a relatively large energy gap between the ground and first excited states. Then as $g$ decreases, the energy gap decreases so that at  some point the energy gap becomes exponentially small. In Fig. \ref{degwf}, we show the corresponding wavefunctions of the ground and first excited states $c(n)$ with different ratios $g/\tau$. It is clear that the ground state does not evolve into cat state until the bimodal system enters the degenerate regime. In the degenerate regime, any small perturbation can mix the two states. As the first excited state is nearly identical with the ground state, but with a $\pi$ phase-shift between the left and right wave-packets, an arbitrary superposition of these two states could easily correspond to a localized state, which is equivalent to a collapse of the Schr\"odinger cat wavefunction. Strictly speaking, this degeneracy implies that to adiabatically create the cat state, the process requires an exponentially long time. Thus experimental realization appears impractical, particularly when considering the inevitable exposure of the system to noise.

\subsection{Dynamic evolution}
\label{dynev}

An alternate approach to a Schr\"odinger cat state, based on dynamical evolution, has been  studied for some time \cite{Micheli2003}. In this subestion, we show that it is feasible to dynamically create the cat state from a coherent state by using a Feshbach resonance to make a sudden change in the scattering length. Our proposed scheme  is as follows. First, our system is prepared in the ground state with repulsive interactions and strong tunnelling between wells. The could be accomplished, e.g. by preparing a BEC in a single well, and then adiabatically raising a barrier to divide the well into two equal parts. Note here, the quantum state of this system is coherent in the sense that each atom in the condensate is independently in a superposition of the two localized states. The interaction parameters in this initial stage are chosen to satisfy $g=\frac{\tau}{0.9N}$ as this will lead to the optimal cat-like state attainable in this scheme \cite{Milburn1997}. Once this state is established, the Feshbach resonance is used to achieve a sudden switch of the sign of the scattering length, i.e. we go from $g=\frac{\tau}{0.9N}$ to $g=-\frac{\tau}{0.9N}$. In order to avoid the collapse of the condensate when scattering length $a$ becomes negative, the amplitude of atomic interaction $g$ should be small.  At this point, with tunneling still on, the initial state is no longer an eigenstate, thus it will start a dynamic evolution under the new Hamilton.

\begin{figure}
\epsfig{figure=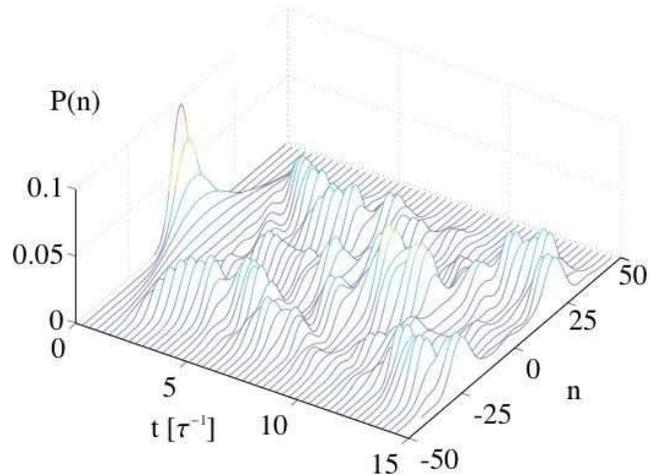, width=9.0cm}
 \caption{Evolution of the probability distribution $P(n)=|c_n|^2$ versus $n$, with $N=100$.  At $t=0$, the initial state is taken as the ground state with$g=\tau/(0.9N)$. Then, the two body interaction is suddenly switched to the negative value $g=-\tau/(0.9N)$. The wavepacket then evolves, where it shows a collapse and revival process. It is seen that, at a time of $t=14.5/\tau$, a well peaked cat state is obtained with two well separated wavepackets centers at around $n=25,75$, respectively. 
 \label{formation-cat}}.
\end{figure}

The evolution of the initial coherent state after the change in the sign of $g$ is shown in Figure \ref{formation-cat}. It is seen that cat-like states are formed periodically, and between two consequent cat states the system approximately revives back to the initial coherent state. We have determined that the best cat state appears at time $t=14.5/\tau$, which corresponds to the final state shown in Fig. \ref{formation-cat}. Once the cat state is formed, one needs to freeze the dynamic evolution at a time when the system is in the cat state. This can be accomplished by suddenly rising the barrier between the two modes to reduce the tunneling coeffient, $\tau$, to zero.

Since the present proposal is based on a dynamic procedure, predicting the time when the cat state is produced $t_{cat}$ is of critical importance. Due to the imposed constraint $\tau=0.9 N g$, which is necessary to obtain an optimal cat-state, 
$t_{cat}$ can be expressed solely in terms of the tunneling rate $\tau$ as
\begin{equation}
	t_{cat}=\frac{\kappa}{\tau},
\end{equation}
where $\kappa$ is determined from our simulations to be $\kappa \approx 14.5$. 

\begin{figure}
\epsfig{figure=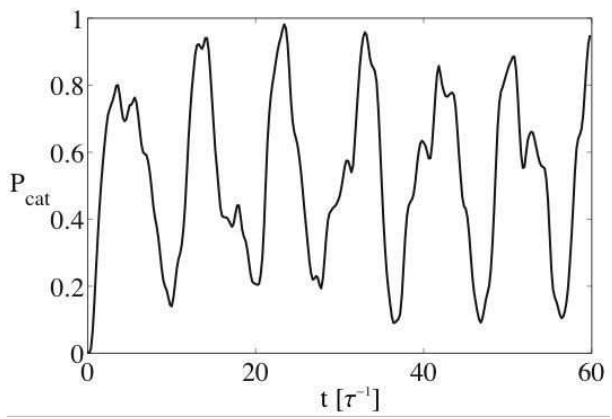, width=8.0cm}
 \caption{Time evolution of the expectation value of the projector $P_{cat}$ with system parameters chosen as in Fig. (\ref{formation-cat}) and a filtering window of width $n_c=15$. A well peaked cat state is created at $t=14.5/\tau, 24/\tau,34/\tau...$..  Note the $P_{cat}$ reaches a minimum at $t\approx10/\tau,20/\tau,30/\tau...$, which corresponds to the revival of the initial coherent state.
 \label{projection}}.
\end{figure}

In Figure \ref{projection}, we show the dynamics evolution of projection operator $P_{cat}$. Again, it is shown that the optimum cat-like state forms at around $t=t_{cat}=14.7\tau$ and reappears again at intervals of $t=10/\tau$. It is also noticed that between each cat state, there are valleys of $P_{cat}$, which correspond to revivals of the initial coherent state.

\begin{figure}
\epsfig{figure=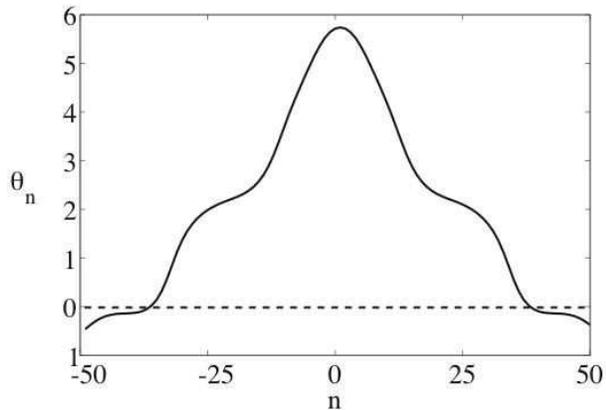, width=8.0cm}
 \caption{Phase distribution of the created Schr\"{o}dinger cat state in Fock space. The $y$ axis is the phase $\theta_n=Arg(c_n)$.
 \label{phase-cat}}.
\end{figure}

It is also of interest to study the relative phases between the Fock states in the cat-like state generated in this manner. 
The symmetry between left and right modes is automatically guaranteed in our model since our system Hamiltonian (\ref{Bose-Hubbard}) and our initial state are both symmetric. In Figure \ref{phase-cat}, we see that the phase of created cat state is exactly symmetric around $n=0$. In addition, we see that the phase flattens somewhat in the vicinity of $n=\pm 25$, where the peaks in the probability are located.

\begin{figure}
\epsfig{figure=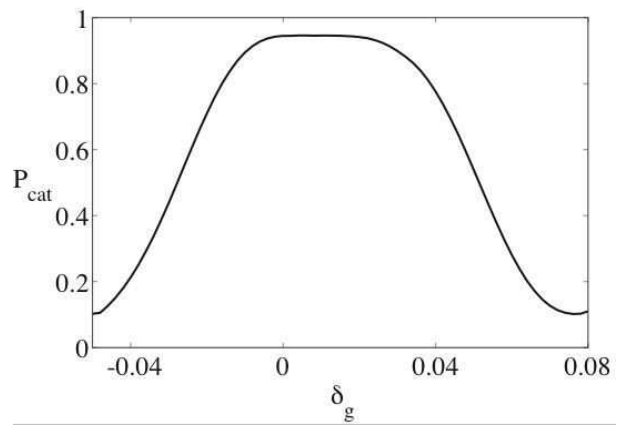, width=8.0cm}
 \caption{The measurement $P_{cat}$ versus interaction strength variation, with $g=\frac{\tau}{0.9N}(1+\delta_g)$. Other system parameters and projection operator is identical to Fig.(\ref{projection}). It is shown that in the vicinity of $\delta_g=0$, the cat state is created. Note that the projection is measured at different times for each $\delta_g$, so that the value of $P_{cat}$ is maximized. \label{projection-measure}}.
\end{figure}

We have claimed that imposing the condition $\tau=0.9Ng$ is necessary to produce the optimal cat state. It is worth pointing out that with other choices, left-right symmetric states can be created, but these are neither well-separated nor double peaked. This is shown in Figure \ref{projection-measure}, where we plot the expectation value of the projector $P_{cat}$ versus $\delta_g$, where we have taken $g=\frac{\tau}{0.9N}(1+\delta_g)$. In this figure, we see the cat-like state is only created in the vicinity of $\delta_g=0$ with an uncertainty of approximately $\pm 5\%$. That means to create a cat state in the present system, a precise control of the atomic interaction $g$ or tunneling rate $\tau$ is required. It is important to note that in generating Figure \ref{projection-measure} we have varied the evolution time $t_{cat}$ for each parameter choice, in order to maximize the projection $P_{cat}$. Thus Figure \ref{projection-measure} can be interpreted as verifying the conditions $g=\frac{\tau}{0.9N}$ and $t_{cat}=14.7/\tau$.

\begin{figure}
\epsfig{figure=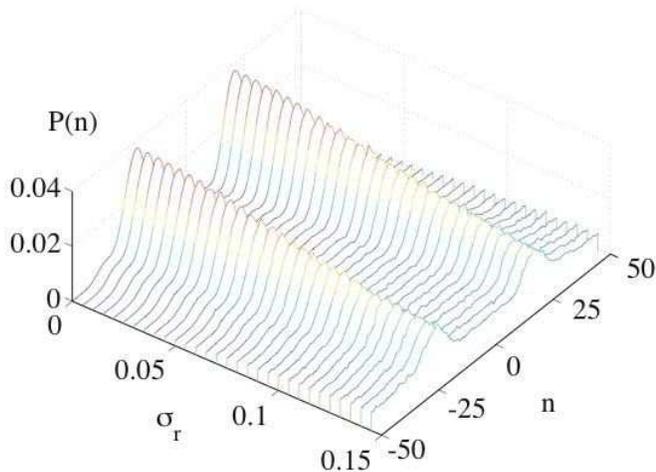, width=9.0cm}
\caption{The probability distribution $P(n)$ at time $t=t_{cat}$ is plotted versus the level of random noise in the initial state $\sigma_r$. We see that even at $\sigma_r=0.05$ the output state is still a good cat-state. This corresponds to random fluctuations in the initial $c_n$'s  at about $5\%$ of their exact ground state values}
\label{stability-cat}.
\end{figure}

Up to now we have only considered the ideal situation where the initial state is exactly prepared in the ground state and the atomic scattering length can be tuned in a precise way. This  may not necessarily hold true under experimental conditions, where random fluctuations in system parameters may affect the formation of  the cat-state. Here, to test the susceptability of the present proposal to such fluctuations, we study the dependence of final cat-state on the precision of the initial state by adding random noise to the exact ground state. Thus, the initial state becomes $c_n+\delta_r$. The random noise, $\delta_r$, is a complex with an amplitude drawn from a Gaussian distribution, and a completely random phase. The probability distributions for the amplitude and phase are therefore 
\begin{eqnarray}
	P(|\delta_r|)&=&\frac{1}{\sqrt{\pi}\sigma_r} e^{-|\delta_r|^2/\sigma^2_r}, \nonumber\\
	P(\theta)&=&\frac{1}{2\pi}.
\end{eqnarray}
Figure \ref{stability-cat} shows the resulting cat-state as a function of $\sigma_r$. We see that a well peaked cat state can be produced up to $\sigma_r\approx 0.05$, which is about $c_0(t=0)/6$, i.e, the peak of initial coherent state we make the cat state from. Thus this method of generating a cat state appears robust against noise in the initial state, which could be do to a finite temperature. 

\begin{figure}
\epsfig{figure=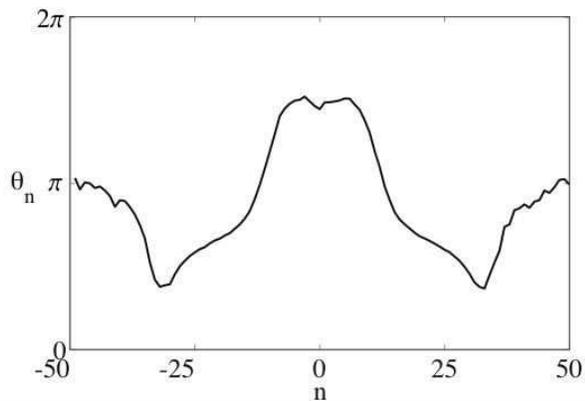, width=8.0cm}
\caption{The phase distribution $\theta_n=\arg(c_n)$ at $t=t_{cat}$ is plotted for the case of random noise in the initial state with $\sigma_r=0.05$. We see that the phase distribution agrees well with Fig. \ref{phase-cat} up to a non-physical overall phase-shift. The disagreement occurs only in the regions where the amplitude of $c_n$ is small.}
\label{randomphasedist}
\end{figure}

\begin{figure}
\epsfig{figure=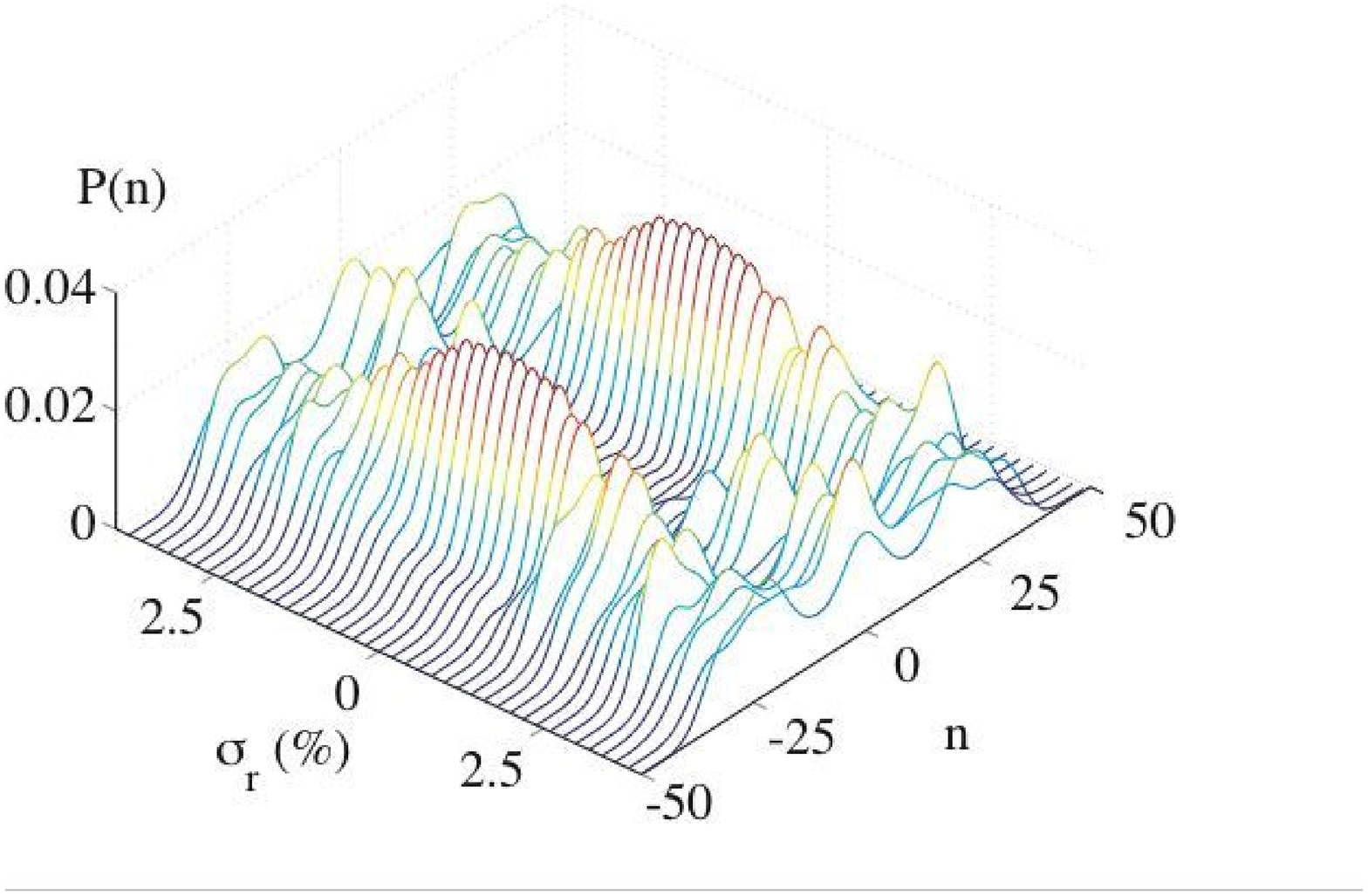, width=9.0cm}
 \caption{ The effect of imprecise control over the  atomic interaction strencth $g$. The noise of $g$ is introduced by adding a real random number, where $\sigma_r$ is the width of the corresponding Gaussian distribution, to the  value of $g$ in Hamiltonian (\ref{Bose-Hubbard}). The cat state is measured at a time of $t=14.5/\tau$ and plotted as a function of $\sigma_r$. It is shown that uncertanties in $g$ up to $3\%$ can be tolerated. \label{randomg}}.
\end{figure}

Another possible source of error in creating the cat state is from the inaccuracy in control of the atomic interaction strength $g$ and/or the tunneling rate $\tau$.  This effect is simulated by adding a Gaussian-distributed random number to negative $g$. We note that this is not equivalent to time-dependent fluctuations in $g$ and $\tau$, but rather represents imprecision in the control over the values of the parameters. The result is shown in Fig. \ref{randomg}, where it is seen that the evolution  system is sensitive to the noise of the atomic interaction $g$, with only about $\pm2\%$ deviation being tolerable. However, it does not necessarily mean that one can not create a cat-state without the well defined $g$. Instead, when $g$ is shifted, the time need to create cat state $t_{cat}$ is also shifted, and one can still obtain the cat-state on another time. The primary difficulty here is that if the variation in $g$ is unknown for a given run, then the precise time to stop the dynamics, $t_{cat}$ is not known a-priori. 

\section{Detection of cat state}
\label{detection}

\begin{figure}
\epsfig{figure=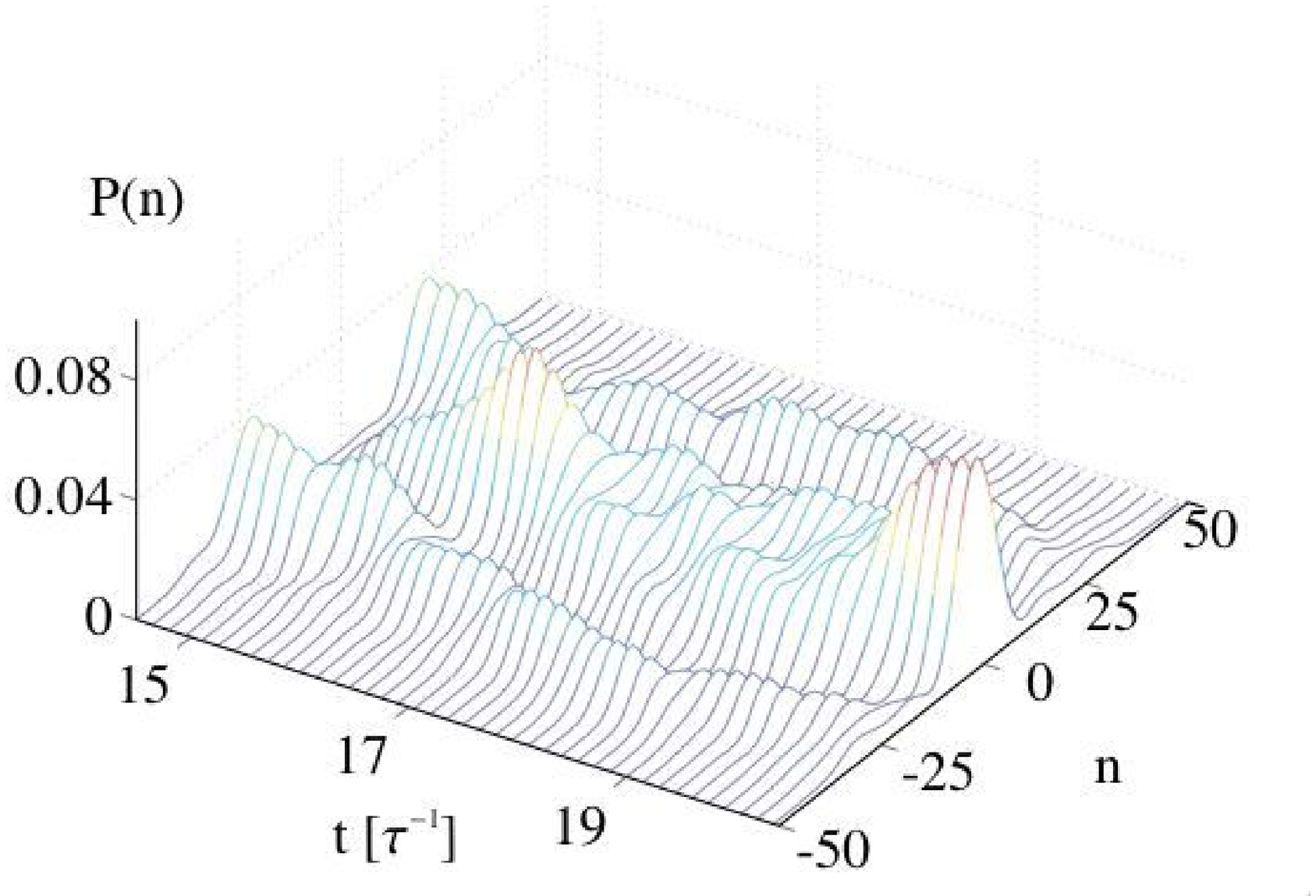, width=9.0cm}
 \caption{Revival dynamics from the cat state back to the coherent state.  This figure shows the continued dynamical evolution following that shown in Fig.\ref{formation-cat} with $g=-0.01$. At $t=14.5/\tau$, the quantum system is in a cat state, then at a time of $t\approx 21$, the initial coherent state is  revived. \label{revival-cat}}.
\end{figure}

Presuming a cat-like state can be created in the above manner, it remains a challenge to experimentally verify the coherence between the macroscopically distinguishable states. Neverthless, a possible method to detect Schr\"{o}dinger cat states in this bimodal BEC's lies in detecting the revival of the initial coherent state. In Fig. \ref{revival-cat} we show the evolution of the cat state starting from $t=t_{cat}$ with the system parameters still satisfying $g=\frac{\tau}{0.9N}$. This shows
that under the condition of no decoherence, the system will evolve back to a coherent state after a certain period of time, which we denote as $t_{rev}$. The revival time $t_{rev}$ again is estimated  from our simulations, yielding $t_{rev}=7/\tau$.

In the presence of decoherence, however, the system may not be exhibit a distinct revival of the coherent state. To illustrate this we first consider two states which we define as the partially incoherent state and  the totally incoherent state.  These states are defined by the introduction of a coherence `length' in Fock-space, $I_{coh}$, so that coherent superpositions
between Fock-states $|n\rangle$ and $|m\rangle$ are destroyed for $|n-m|\gg I_{coh}$, while superpositions satisfying $|n-m|\lesssim I_{coh}$ are relatively unaffected by decoherence.
The partially incoherent state is then defined in the sense that the coherence length $I_{coh}$ is small compared to the distance between the two separated peaks of the cat state, but larger than the width of either wavepacket, i.e., $I_{coh} < n_0$ yet $I_{coh} > \sigma$  in Eq. (\ref{cat-Gaussian}). For the completely incoherent state, on the other hand, the cat state is taken to have collapsed to a single Fock state, i.e. a mixed state maintaining the probability distribution of the cat-state but with to coherence between Fock-states. 

\begin{figure}
\epsfig{figure=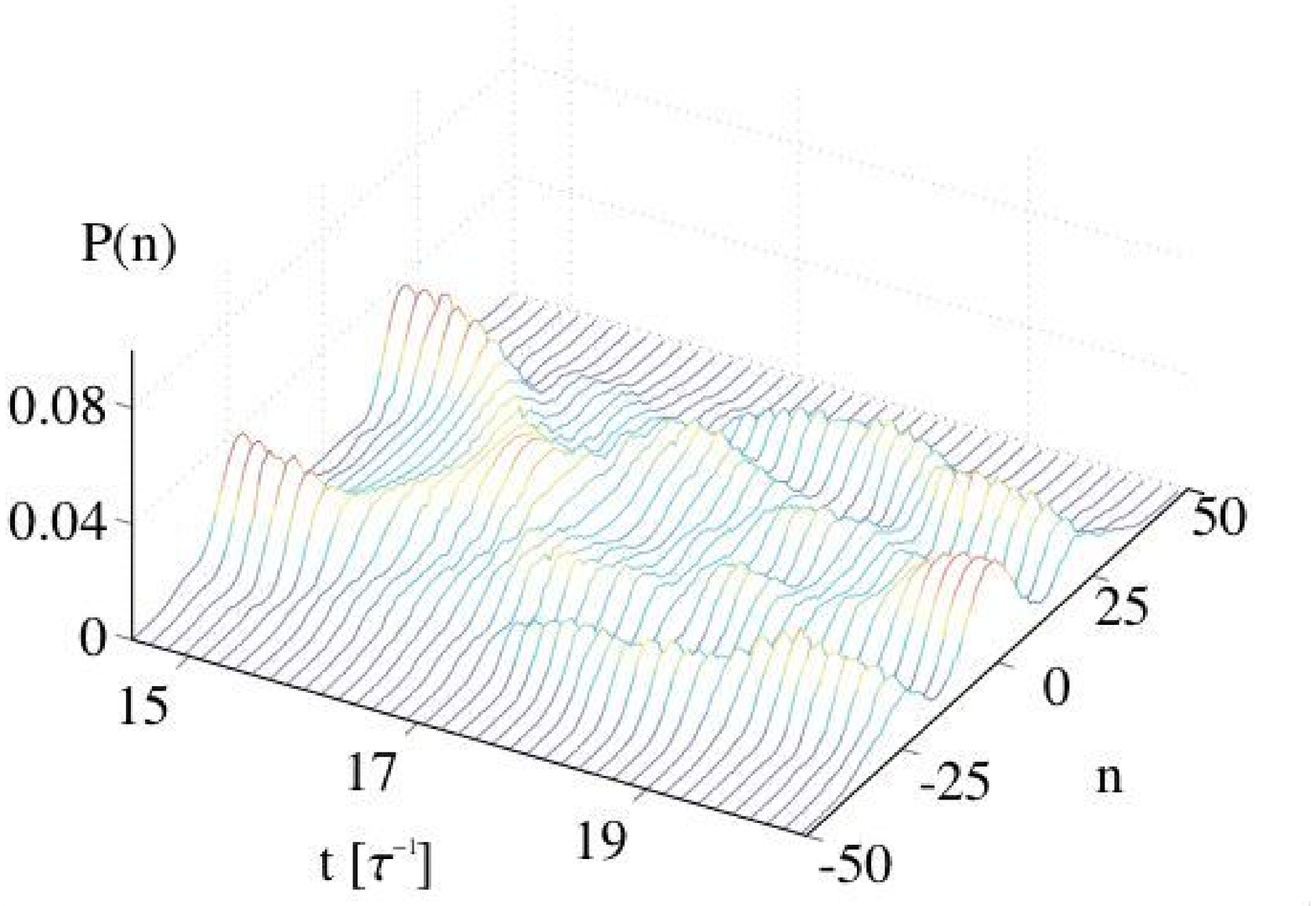, width=9.0cm}
 \caption{Dynamical evolution of a partially incoherent cat state. The parameters are chosen the same as Fig.\ref{revival-cat}. At $t=14.5/\tau$, the quantum system is in a mixed state with no coherence between the two wavepackets. Then as it evolves, the system exhibits oscillations, but will not evolve back to the initial state.    \label{partial-decoherence}}.
\end{figure}

\begin{figure}
\epsfig{figure=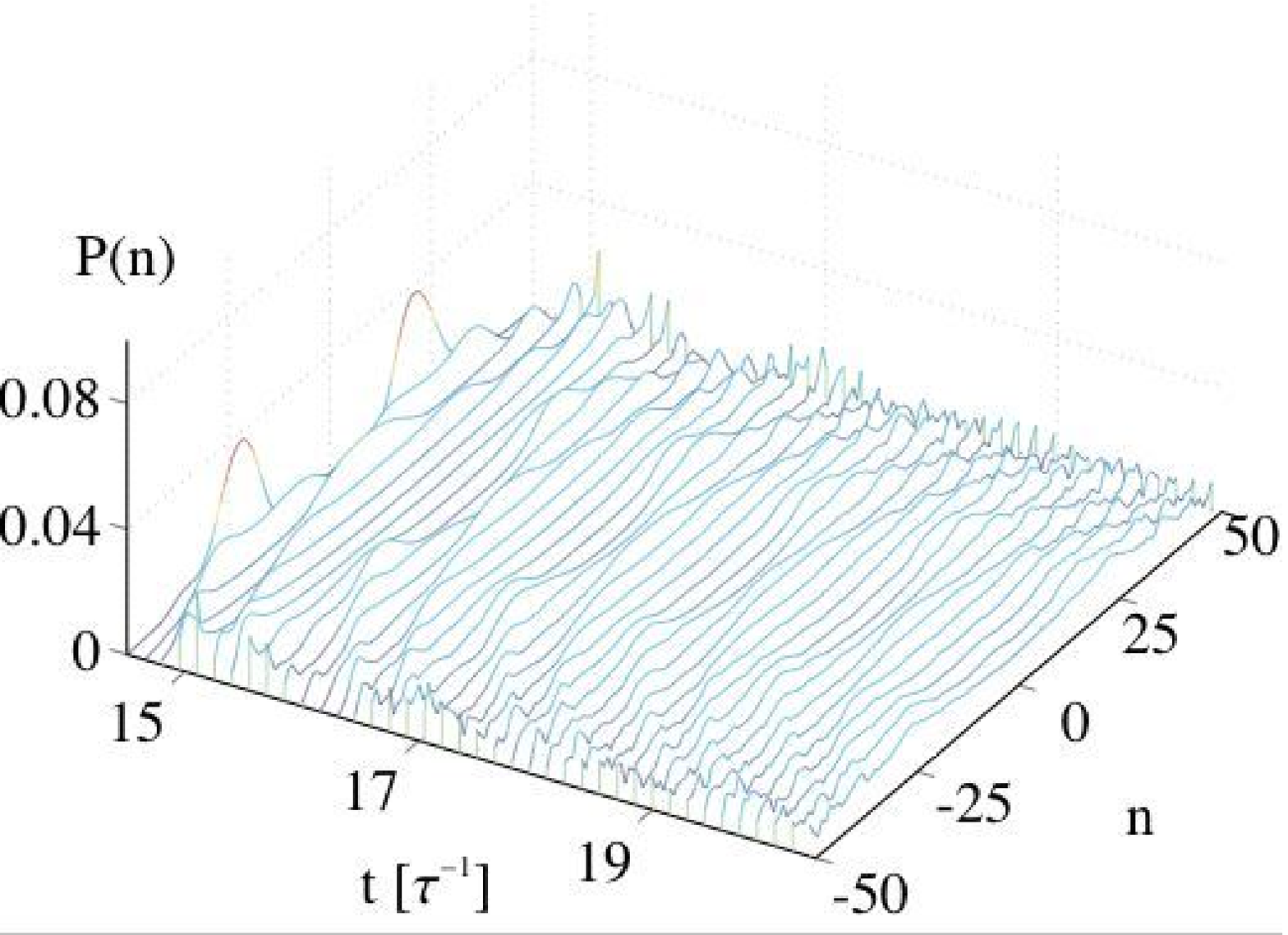, width=9.0cm}
 \caption{Dynamic evolution of completely decoherent cat state. At $t=14.5/\tau$, we let the system collapse onto a statistical mixture of Fock-states. It is seen that the dynamic is completely disordered.
  \label{complete-decoherence}}.
\end{figure}

The evolution of those two states in time is shown in Fig. \ref{partial-decoherence} and Fig.\ref{complete-decoherence}, respectively. Unlike the cat state, the incoherent states will not exhibit a revival of the initial coherent state at $t=t_{rev}$. The partially incoherent state evolves instead into a disinct three-peaked state, whereas the completely incoherent state exhibits small oscillations with no discernable peaks after a very short collapse time. 

Based on these considerations, one prospective method to  verify the coherence of the cat state would involve: i) first, measuring the probability distribution, $P(n)$, which is accomplished by repeated measurements of the atom-number in each mode; ii) then after forming the cat-state, the state should be held at $\tau=0$ for a time $t=t_{hold}$ and then with tunneling restored, the revival of the initial coherent state should be observed after a time $t_{rev}$;  iii)  a known source of decoherence should be added while holding the system at $\tau=0$ for duration $t_{hold}$, after which a lack of revival should be observed. To add decoherence, one could simply add a laser field to the system. This will introduce decoherence due to spontaneous photon scattering, as described in Section \ref{decoherence}. If the revival of the initial state can be observed in the case of no additional decoherence, but disappears when the decoherence is added in a controlled way, than one should have a strong claim for the observation of coherence between macrscopically distinguishable states.

\section{Laser-induced decoherence and losses}
\label{decoherence}

While trapped ultracold atoms are typically well-isolated from the environment, the presence of lasers for trapping and/or detection of atoms will polarize the atoms and thus couple them to the vacuum modes of the electromagnetic field. This coupling allows for spontaneous Rayleigh scattering and/or diffraction of laser photons, which can lead to losses and/or decoherence of the atomic system. 
If the scattered light field acquires sufficient information to determine the atom-number difference between the left and right wells, then this spontaneous scattering will lead to a dynamical collapse from the cat-state into a mixture of Fock states, an example of environmentally-induced decoherence.

In addition, the condensates are in contact with a thermal cloud on non-condensed atoms. In \cite{Diego2000},  it was determined that such a coupling will lead to a rapid collapse of the cat-state. We note, however, that the treatment in \cite{Diego2000} made the approximation that 
correlations between the condensate and the cloud decay much faster than the time-scale of the condensate dynamics, i.e. the thermal cloud was treated as a Markovian reservoir. At present we question the validity of the Markov approximation for such a system. The coherence time can be quite long because (i) the cloud temperature is very low and the cloud is very dilute (so that collisions between cloud atoms are rare), and (ii) the cloud is confined, so that information does not propagate away from the condensate region. A non-Markovian treatment of thermal-cloud-induced decoherence is beyond the scope of this work. At present, we focus primarily on the decoherence induced by coupling to the electromagnetic field.

The structure of this section is as follows: in subsection A, we will first investigate the decoherence mechanics of condensates induced by laser scattering, where the general master equation is derived. Then, in subsection B, we discuss the condensate losses due to the inelastic collision, and the loss rate is obtained for one-mode condensate. In subsection C, we show the decoherence of the two-mode cat states in the elastic scattering regime, defined as the regime where the photon-recoil is not sufficiently strong to remove the atom from its initial center-of-mass state. The system dynamics under this loss-less decoherence is studied in subsection D.. Lastly, the atom-loss and decoherence of cat-states in the inelastic regime is studied in subsection E,  where we find that the cat-states created via dynamic evolution are less susceptible to decoherence-induced collapse than the extreme cat-state, i.e. the true ground-state of the system, which we have argued is inaccessible via adiabatic evolution.

\subsection{Master equation}

If there is an optical component to the trapping potential, the atoms in the BEC will have an induced electric dipole-moment which oscillates at the laser frequency. This dipole-moment will couple the atoms to the electromagnetic vacuum, resulting in spontaneous Rayleigh scattering and/or diffraction of the laser beam. In addition, there will be laser-induced many-atom effects, primarily dipole-dipole interactions and collective effects such as superradiance. As we will see, these effects will lead to decoherence and collapse of the cat-states, even in the absence of recoil-induced losses from the two cat modes. This decoherence is associated with the possibility for scattered and/or diffracted photons to carry away information sufficient to reveal the distribution of atoms between the two modes. 

In order to model these effects we need to derive the master-equation which governs the density operator of the atomic field.
In the presence of laser coupling to a BEC system, the total Hamiltonian for the system, which includes  the Hamiltonian for the atoms in condensate $H_s$, the reservoir of vacuum electromagnetic field $H_s$ and the interaction between them in the dipole approximation $V_{sr}$ can be written as
\begin{equation} 
\label{atom-laser}
	H=H_s+H_r+V_{sr}, 
\end{equation}
\begin{equation}
\label{Hs}
	H_s=\int d^3r \Delta\hat{\Psi}^\dagger_e(\mathbf{r})\hat{\Psi}_e(\mathbf{r})+\int d^3r \Omega(\mathbf{r})
	 \hat{\Psi}^\dagger_e(\mathbf{r})\hat{\Psi}_g(\mathbf{r}), 
\end{equation}
\begin{equation}
\label{Hr}
	H_r = \sum_{k\lambda}\omega_k\hat{a}^{\dagger}_{\bfk\lambda}\hat{a}_{\bfk\lambda}, 
\end{equation}
\begin{equation}
\label{Vsr}
	V_{sr} = \sum_{k\lambda} \int d^3r \,g_{k\lambda} e^{i\mathbf{k}\cdot\mathbf{r}}e^{i\omega_L
	t}\hat{\Psi}^\dagger_e(\mathbf{r})\hat{\Psi}_g(\mathbf{r})\hat{a}_{\bfk\lambda}+H.c.,
\end{equation}
where $\hat{\Psi}_e(\mathbf{r})$ and $\hat{\Psi}_g(\mathbf{r})$ are the annihilation operators for atoms in the excited and ground state respectively, and  $\hat{a}_{\bfk\lambda}$ is the photon annihilation operator for momentum $\hbar\bfk$ and polarization state $\lambda$. In addition,  $\omega_e$, $\omega_L$ and $\omega_k$ denote the frequency of atomic excited state, laser field, and the reservoir photons respectively,  while $\Delta=\omega_e-\omega_L$ is the detunning of laser field from the excited state. The atom-laser interaction is governed by the  Rabi-frequency $\Omega(\bfr)$, while the atom-reservoir interaction is governed by $g_{k\lambda}=\sqrt{\frac{\omega_k}{2\hbar\epsilon_0V_e}}\mathbf{d}\cdot\mathbf{\epsilon}_{k\lambda}$, with $\mathbf{d}$ being the atomic dipole moment in mks units. We note that for the purpose of deriving the master equation it is not necessary to include terms in $H_s$ governing the free evolution of the ground state, as this involves very slow time scales relative to the dynamics of the excited-state/EM-vacuum system. These terms can be added to the system Hamiltonian at the end of the calculation, as any energy-shifts they introduce will be negligible compared to the natural line-width of the excited state.

The quantum state of the system+reservoir is described by the density operator $\rho_{sr}$.
An effective equation of motion for the reduced system density operator, $\rho_s=tr_r\{\rho_{sr}\}$, can be derived via second-order perturbation theory 
by following the standard approach (see for example \cite{meystrebook}). For the case of a zero-temperature bath this leads directly to the master equation
\begin{eqnarray}
\label{reduced-master}
	\dot\rho_s(t)&=&-iH_s\rho_s(t)\nonumber\\
	&-&\int d^3r d^3r' \,\left[ L(\mathbf{r}-
	 \mathbf{r}')\hat{\Psi}^\dagger_e(\mathbf{r})\hat{\Psi}_g(\mathbf{r})\hat{\Psi}^\dagger_g(\mathbf{r'})
	 \hat{\Psi}_e(\mathbf{r}')\rho_s(t)\right. \nonumber\\
	 &-& \left. L^\ast(\mathbf{r}-\mathbf{r}')\hat{\Psi}^\dagger_g(\mathbf{r})
          \hat{\Psi}_e(\mathbf{r})\rho_s(t)\hat{\Psi}^\dagger_e(\mathbf{r}') \hat{\Psi}_g(\mathbf{r'})\right]\nonumber\\
          &+&H.c.,
\end{eqnarray}
where 
\begin{equation} 
\label{L-master}
 	L(\mathbf{r}-\mathbf{r}')=\sum_{k\lambda} |g_{k\lambda}|^2 e^{i\mathbf{k}\cdot(\mathbf{r}-
	\mathbf{r}')}\{\pi\delta(\omega_k-	
	\omega_L)+\frac{i\mathcal{P}}{\omega_k-\omega_L}\},
 \end{equation}
with $\mathcal{P}$ indicating a principal value. In the limit of infinite quantization volume, the summation becomes an integral, which can be done analytically \cite{Zhang1994}, yielding
\begin{equation}
	L(\bfr-\bfr')=\frac{3\Gamma}{4}e^{-i\zeta}\left[\sin^2\theta\frac{i}{\zeta}+ (1-3\cos^2\theta)\left[\frac{1}{\zeta^2}-\frac{i}{\zeta^2}\right]\right]
\end{equation}
where  $\Gamma=\omega^3_L d^2/(3\pi\epsilon_0\hbar c^3)$ is the single-atom spontaneous emission rate, $\zeta=k_L |\mathbf{r}-\mathbf{r}'|$ and $\theta$ is the angle between the dipole moment $\mathbf{d}$ and $\mathbf{r}-\mathbf{r}'$. The real part of $L(\bfr-\bfr')$ will contribute to decoherence in the master equation (\ref{reduced-master}), while the imaginary part contributes an energy shift due to photon exchange between atoms. 
In the limit $|\bfr-\bfr'|\to0$ we find
\begin{equation}
L(0)=\frac{\Gamma}{2}+i\delta,
\end{equation}
where the imaginary part, $\delta$ is an infinite quantity, reflecting the fact that the standard two-level atom-field interaction model is not properly renomalized. Physically, this term is a Lamb-type shift in the energy of the excited-state due to interaction with the vacuum modes of the EM field. In our system we can always choose a rotating frame to absorb this shift, so that we can take $\delta=0$ without loss of generality.   

For most optical traps, the laser is detuned very far from the atomic resonance frequency, so that the excited state occupation number is $\ll 1$. In this regime it is useful to perform an adiabatic elimination of the excited-state operators. The resulting master equation, involving only ground-state operators, is
\begin{widetext} 
\begin{eqnarray}
\label{mastereq}
	\dot\rho_s(t) &=& -iH_s\rho_s(t)-i\int d^3r\, \frac{|\Omega(\bfr)|^2}{\Delta}\hPsi^\dag(\bfr)\hPsi(\bfr)
        - \int d^3r\, d^3r'\Omega(\bfr) \frac{L(\bfr-\bfr')}{\Delta^2}\Omega^\ast(\bfr')
	\hat{\Psi}^\dagger(\mathbf{r})\hat{\Psi}(\mathbf{r})
	\hat{\Psi}^\dag(\mathbf{r}')\hat{\Psi}(\mathbf{r}')\rho_s(t)\nonumber \\  
	& &+  \int d^3r d^3r' \Omega^\ast(\bfr)\frac{L^\ast(\mathbf{r}- \mathbf{r}')}{\Delta^2}\Omega(\bfr')	\hat{\Psi}^\dagger(\mathbf{r})\hat{\Psi}(\mathbf{r})\rho_s(t)\hat{\Psi}^\dagger(\mathbf{r'})\hat{\Psi}(\mathbf{r}')
	+ H.c.,
\end{eqnarray}
\end{widetext}
where $\hPsi(\bfr)\equiv\hPsi_g(\bfr)$.
The details of this derivation are presented in Appendix \ref{adelim}.
  
 \subsection{Condensate losses: elastic and inelastic regimes}
 \label{condensate-loss}
We will now employ this master equation to describe decoherence and/or atom-loss in our two-mode atomic system.
Underlying the dynamics governed by master equation (\ref{mastereq}) is the exchange of photons between atoms and between an atom and the EM vacuum. As such photon exchanges can involve significant momentum exchanges due to photon recoil, it is useful to first define  elastic and inelastic regimes with respect to the initial atomic center-of-mass state. If the spatial size of the initial atomic mode is small compared to the laser wavelength, then one would be in the elastic regime, as photon-recoil would not be sufficient to carry the atom outside of the momentum distribution of the initial state. In the inelastic regime, on the other hand, the initial mode is large compared to the laser wavelength, which implies that it is narrow in momentum space compared to the photon momentum, so that the recoil from a single photon is sufficient to remove an atom from its initial mode and place it in an orthogonal mode.

In the strongly elastic regime, there will be negligible loss of atoms from their initial modes. As we shall demonstrate, the master equation (\ref{mastereq}) may still lead to decoherence/collapse of the two-mode cat-states. In the inelastic regime, on the other hand, the predominant effect of the atom-field interaction will be scattering of atoms out of the initial modes and into a quasi-continuum of modes. This loss will be accompanied by a collapse of state of the remaining atoms into a mixture of states, which may or may not be cat-like in themselves. The details of this decoherence process are given in subsection \ref{inelastic}.

\begin{figure}
\epsfig{figure=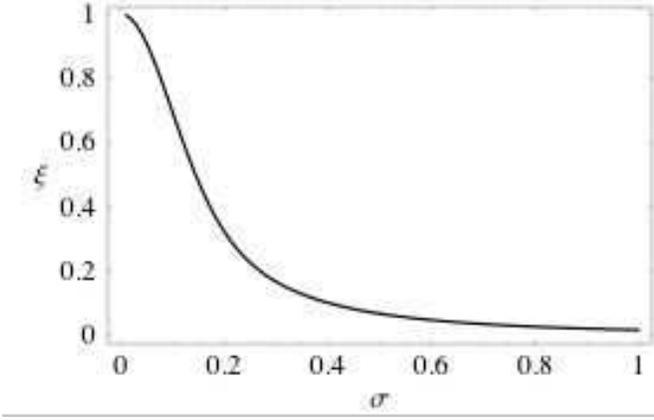, width=9.0cm}
\caption{\label{fig-inelastic} Dependence of the inelastic loss parameter $\xi$ on the spatial size of condensate, which is indicated by the dimensionless parameter $\sigma$ as shown in Eq. (\ref{Gausswf}). It is seen that $\xi$ is approximately unity when the size of condensate is much less than the laser wavelength, i.e. $\sigma\ll1$, which means there is negligible atom loss. As the condensate size approaches half $\lambda_L$, $\xi$ decreases to around $0$, corresponding to the maximum loss rate.}
\end{figure}

To derive the atom decay rate of a single atomic-field mode, we will need to make use of the commutation relation  
\begin{equation} 
	[\hat{c}^\dagger \hat{c},\hat{\Psi}^\dagger(\mathbf{r})\hat{\Psi}(\mathbf{r})]=-
	\phi(\bfr)\hat{c}^\dag\hPsi(\bfr)-\phi^\ast(\bfr)\hPsi^\dag(\bfr)\hat{c},
\end{equation}
where $\hat{c}$ is the annihilation operator for atoms in mode $\phi(\bfr)$. Then, from the master equation (\ref{mastereq}), it is straight forward to obtain the evolution of the mean mode occupation $n_0=\langle\hat{c}^\dagger\hat{c}\rangle$ as
\begin{eqnarray}
\label{dndt}
	\dot n_0 &=&-\int d^3rd^3r'\, \frac{G(\bfr,\bfr')}{\Delta^2}\left[\phi^\ast(\bfr)\delta^3(\bfr-\bfr')
	\langle\hat{c}^\dag\hPsi(\bfr')\rangle\right.\nonumber\\
	&-&\phi(\bfr)\phi^\ast(\bfr')\langle\hPsi^\dag(\bfr)\hPsi(\bfr')\rangle
	+ \phi^\ast(\bfr)\langle\hat{c}^\dag\hPsi^\dag(\bfr')\hPsi(\bfr)\hPsi(\bfr')\rangle\nonumber\\
	&-&\left. \phi(\bfr)\langle\hPsi^\dag(\bfr)\hPsi^\dag(\bfr')\hPsi(\bfr')\hat{c}\rangle\right],	
\end{eqnarray}
where $G(\bfr,\bfr')=\Omega(\bfr)L(\bfr-\bfr')\Omega^\ast(\bfr')$.
Making a single mode approximation,
\begin{equation}
	\hat{\Psi}(\mathbf{r})=\phi(\mathbf{r})\hat{c},
\end{equation}
one obtains
\begin{eqnarray}
\label{dndtsm}
	\dot{n}_0&=&-\frac{2}{\Delta^2}\int d^3rd^3r'\, \Re\left[G(\bfr,\bfr')\right]|\phi(\bfr)|^2\nonumber\\
	& &\qquad\qquad\qquad \times \left[\delta^3(\bfr-\bfr')-|\phi(\bfr')|^2\right]n_0.
\end{eqnarray}
To better understand the physics described by this equation, we make the simplifying assumption $\Omega(\bfr)\approx\Omega e^{i\bfk_L\cdot(\bfr-\bfr')}$, where $\bfk_L$ is the laser wave-vector, satisfying $\bfk_L\cdot{\bf d}=0$. This leads to
\begin{equation}
\label{dndtsimple}
	\dot n_0 = -\frac{|\Omega|^2}{\Delta^2}\Gamma(1-\xi)n_0,
\end{equation}	
where
\begin{equation}
\label{xi}	
       \xi = \frac{2}{\Gamma}\int d^3r d^3r' 
      \Re[e^{i\bfk_L\cdot(\bfr-\bfr')}L(\mathbf{r}-\mathbf{r}')]|\phi_g(\mathbf{r})|^2|\phi_g(\mathbf{r}')|^2.
\end{equation}
In the elastic regime we have $|\phi(\bfr)|^2\approx\delta(\bfr)$ relative to $L(\bfr-\bfr')$, which leads to $\xi=1$ so that there is no atomic loss, as expected. As the size of condensate grows, $\xi$ will decrease, as shown in Fig. \ref{fig-inelastic}, which plots $\xi$ versus condensate size for the case of a spherically symmetric Gaussian condensate wavefuction
\begin{equation}
\label{Gausswf}
 	\phi(\mathbf{r})=\frac{1}{(\sigma \lambda_L)^{3/2}\pi^{3/4}}e^{-r^2/(2\sigma^2\lambda_L^2)},
 \end{equation}
where $\sigma$ is the ratio of the condensate size to the laser wavelength, $\lambda_L$.
From the figure, we see that the decay rate of the condensate can be neglected under the condition $\sigma\ll 1$, while for a mode whose size is comparable to or larger than the laser wavelength, we have $\xi\to 0$ so that the population will decay at the expected rate $|\Omega|^2\Gamma/\Delta^2$, which is simply the electronically-excited state fraction times the excited state lifetime.

\subsection{Decoherence of two-mode cat states in elastic regime}
\label{decoherence-delta}

In the present scheme, $N$ atoms are trapped in a double-well potential with a finite spatial separation between the wells.  Coupling of this system to the EM vacuum modes will result in decoherence only when scattered photons carry away sufficient information about the number of atoms in each well. It thus follows that if the separation between the two condensates, $s$, is less than the laser wavelength $\lambda_L$, no decoherence should occur, as the information contained in the photons is diffraction-limited to a resolution of no better than a wavelength. On the other hand, if $s$ is larger than $\lambda_L$, the locations of the atoms can in-principle be determined from the phase-information contained in the scattered photons, this 
irreversible transfer of information to the environment will be reflected in decoherence of fock-space-superpositions of different atom-number distributions.

In the deeply elastic regime, the size of the left and right modes is the smallest length scale in the problem, so that we can replace the atomic density distributions with delta-functions, or equivalently we take
\begin{equation}
\label{elasticmodes}
	\hat{\Psi}(\mathbf{r})=\sqrt{\delta(\mathbf{r}-\frac{\mathbf{s}}{2})}\hat{c}_L+\sqrt{\delta(\mathbf{r}+\frac{\mathbf{s}}{2})}\hat{c}_R,
\end{equation} 
where $\hat{c}_L$ and $\hat{c}_R$ are the annihilation operators for the left and right mode respectively. 
Inserting this expansion into the master equation (\ref{mastereq}) and assuming again for simplicity $\Omega(\bfr)=\Omega e^{i\bfk_L\cdot\bfr}$, gives
\begin{eqnarray}
\label{two-mode-master-delta}
	\dot\rho_s &=& -\frac{|\Omega|^2\Gamma}{2\Delta^2}
	\left[\hat{c}^\dagger_L\hat{c}_L\hat{c}^\dagger_L\hat{c}_L\rho_s
	+\hat{c}^\dagger_R\hat{c}_R\hat{c}^\dagger_R\hat{c}_R\rho_s\right.\nonumber\\
	& &+2\cos(\bfk_L\cdot{\bf s})\mu(\mathbf{s}) \hat{c}^\dagger_L \hat{c}_L \hat{c}^\dagger_R\hat{c}_R\rho_s
   	-\hat{c}^\dagger_L\hat{c}_L\rho_s \hat{c}^\dagger_L\hat{c}_L\nonumber\\
	& &-\hat{c}^\dagger_R\hat{c}_R\rho_s\hat{c}^\dagger_R\hat{c}_R 
	-\cos(\bfk_L\cdot{\bf s})\mu(\mathbf{s})\hat{c}^\dagger_L \hat{c}_L \rho_s \hat{c}^\dagger_R\hat{c}_R\nonumber\\
	& &-\left. \cos(\bfk_L\cdot{\bf s})\mu(\mathbf{s})\hat{c}^\dagger_R \hat{c}_R \rho_s \hat{c}^\dagger_L \hat{c}_L)\right]+H.c 
\end{eqnarray}
where again the imaginary parts are absorbed via a transformation to the appropriate rotating frame. The resulting decoherence is governed by the parameter $\mu(\mathbf{s})$, defined as
\begin{eqnarray}
	\mu(\mathbf{s}) & \equiv&  2\Re[L(\mathbf{s})]/\Gamma \nonumber \\
	&=&\frac{3}{2}[\sin^2\theta_\mathbf{s} \frac{\sin \zeta}{\zeta}+(1-3\cos^2\theta_\mathbf{s})( \frac{\cos \zeta}{\zeta^2}-\frac{\sin \zeta}{\zeta^3})].  \nonumber\\
\end{eqnarray}
where $\zeta=2\pi s/\lambda_L$ and $\theta_\mathbf{s}$ is the angle between the induced polarization direction $\mathbf{d}$ and the relative coordinate $\mathbf{s}$.  

From this master equation (\ref{two-mode-master-delta}), we  derive the equation of motion for the fock-space matrix elements $\rho_{nm}=\langle n|\rho_s|m\rangle$, yielding
\begin{equation}
\label{meqtd}
	\dot \rho_{nm}=- \Gamma\frac{|\Omega|^2}{\Delta^2}(n-m)^2(1-\cos(\bfk_L\cdot{\bf s})\mu(\mathbf{s}))\rho_{nm}.
\end{equation}
This shows that the diagonal matrix elements (m=n) do not decay, reflecting the absence of condensate losses in the elastic regime. 
The off-diagonal elements, on the other hand, decay as at a rate proportional to $(1-\cos(\bfk_L\cdot{\bf s})\mu(\mathbf{s})) (n-m)^2$. 

\begin{figure}
\epsfig{figure=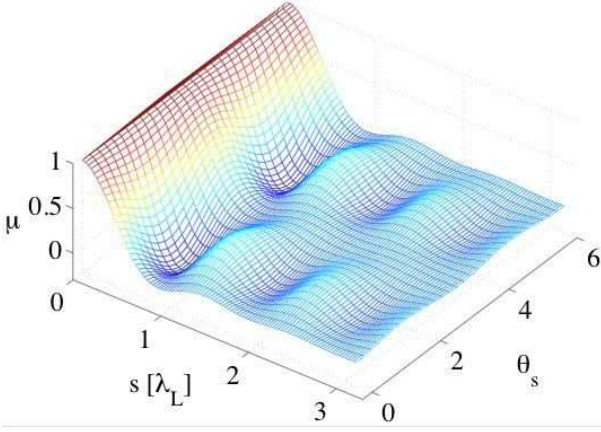, width=8.0cm}
 \caption{The decoherence parameter $\mu(\mathbf{s})$ as a function of $\theta_\mathbf{s}$ and $s/\lambda_L$. From the diagram, $\mu(\mathbf{s})$ is approximately $1$ when the separation of two modes $s$ is much smaller than the laser wavelength $\lambda_L$. As $s$ increases, $\mu(\mathbf{s})$ decreases at a speed of $1/s$ and eventually approaches zero as $s$ goes much larger than $\lambda_L$. Note that $\mu(\mathbf{s})$ is also dependent on $\theta_\mathbf{s}$, where at large $s$, $\mu(\mathbf{s})\sim \sin^2\theta_\mathbf{s}$.    
 \label{fig-mu}}.
\end{figure}

The dependence of $\mu(\mathbf{s})$ on the mode separation is shown in Fig. \ref{fig-mu}, which plots $\mu(\mathbf{s})$ versus $s/\lambda_L$ and $\theta_\mathbf{s}=\arg {\bf s}$. We see that under the condition of $s\ll \lambda_L$ we have $\cos(\bfk\cdot{\bf s})\mu(\mathbf{s})\approx 1$, and therefore no decoherence. As $s$ increases,  $\mu(\mathbf{s})$ will decrease and the decoherence rate will increase. In the limit of $s \gg \lambda_L$, we have $\mu(\mathbf{s}) \approx 0$, and the decoherence converges to the rate $\gamma_{nm}=\frac{|\Omega|^2}{\Delta^2}\Gamma(n-m)^2$.  It is interesting to note that $\mu(\mathbf{s})$ is angle-dependent, as seen  in Fig.\ref{fig-mu}. 

To understand the reason for this decoherence rate(\ref{meqtd}) we will consider the limiting cases $s\gg\lambda_L$ and $s\ll\lambda_L$, corresponding to a mode separation much larger than or much smaller than the laser wavelength, respectively.
For the case where the mode separation is large compared to the laser wavelength, we have $\mu(\mathbf{s})\to 0$ so that the decoherence rate is $\gamma_{nm}=\frac{|\Omega|^2}{\Delta^2}\Gamma (n-m)^2$. 
Our goal is thus to explain the peculiar $(n-m)^2$ dependence in this expression. We begin by noting that in this regime the phase information in the scattered photon wavefronts can in principle determine whether the photon scattered from the L or R mode. Thus the question becomes: on what timescale, $T$, does the radiation field scattered by L or R acquire sufficient information to distinguish the state $|n\rangle$ from the state $|m\rangle$. This timescale will then determine the decay rate for the off-diagonal elements $\rho_{nm}$, which give the degree of coherence between the states $|n\rangle$ and $m\rangle$. The rate at which photons will scatter from the left mode is $\frac{|\Omega|^2}{\Delta^2}\Gamma n_L^2$, where $n_L$ is the number of atoms in the left mode. The reason for the $n_L^2$ dependence, rather than the usual $n_L$, is due to Bose stimulation, as the atoms remain in their initial mode after elastic scattering. 

The actual scattering of photons is a random process, so that the number of L scattered photons $n_p$ over a time interval $t$ is drawn from a distribution function $P_L(n_p,n_L,t)$. It is reasonable to assume that $P_L(n_p,n_L,t)$ is at least approximately Poissonian, with a center at $\bar{n}_p(n_L,t)=\frac{|\Omega|^2}{\Delta^2}\Gamma n_L^2t$, and a width $\Delta n_p(n_L,t)=\sqrt{\bar{n}_p(n_L,t)}=\frac{|\Omega|}{\Delta}\sqrt{\Gamma t}\, n_L$. Based on this assumption, the criterion for distinguishing state $|n\rangle$ from state $|m\rangle$ is that the distributions $p_L(n_p,n_L,t)$ and $p_L(n_p,m_L,t)$ should be distinguishable, so that the scattered photon number can be attributed to one distribution or the other. In this expression $n_L$ is the number of atoms in mode L for state $|n\rangle$ and  This requires minimal overlap between the two distributions, which can be approximately formulated as 
\begin{equation}
\label{overlap}
	\bar{n}_p(n_L,T)+\Delta n_p(n_L,T)=\bar{n}_p(m_L,T)-\Delta n_p(m_L,T),
\end{equation}
where we have temporarily assumed $m_L>n_L$.
Solving this equation for the decoherence rate $\gamma_{nm}=1/t_d$ yields
\begin{equation}
\label{gammanm1}
	\gamma_{nm}=\frac{|\Omega|^2}{\Delta^2}\Gamma \frac{\left[m_L^2-n_L^2\right]^2}{\left[m_L+n_L\right]^2}.
\end{equation}
From eq. (\ref{Nn}) it follows that
\begin{eqnarray}
\label{nL}
	n_L&=&\frac{N}{2}-n\nonumber\\
	n_R&=&\frac{N}{2}+n,
\end{eqnarray}
which also hold for $m_L$ and $m_R$ with a simple substitution $n\to m$.
Inserting this into (\ref{gammanm1}) yields
\begin{equation}
\label{gammanm2}
	\gamma_{nm}=\frac{|\Omega|^2}{\Delta^2}\Gamma(n-m)^2.
\end{equation}	
We note that the result for mode R instead of L can be found via $n\leftrightarrow -n$ and $m\leftrightarrow -m$, while the result for $m_L<n_L$ can be found  via $n\leftrightarrow m$, all of which leave the result (\ref{gammanm2}) unchanged. Thus we have verified our interpretation that the decay of the $\rho_nm$ coherence is governed by the timescale on which the information carried by the scattered light becomes sufficient to distinguish between the states $|n\rangle$ and $|m\rangle$.

If we consider the opposite case $s\ll\lambda_L$ we can no longer assume that scattered photons carry information concerning which mode they scattered from, due to the standard diffraction limit. In this case one might be tempted to assume that the total scattering rate is proportional to $n_L^2 +n_R^2=N^2/2+2n^2$, as photons must still scatter from one mode or the other and each mode will experience Bose-stimulation of the recoiling atom. This would imply that the total scattering rate would be different for the states $|n\rangle$ and $|m\rangle$, so that given sufficient time the environment could learn which state is scattering the light and thus destroy the coherence. This effect, however, is compensated for by superradiance \cite{dicke} so that the scattering rate is proportional simply to $N^2$, which is the same for both $|n\rangle$ and $|m\rangle$. Superradiance is a many-body quantum-interference effect whereby atoms within a distance less than the optical wavelength experience enhanced decay, identical to Bose stimulation, even for distinguishable atoms.

To verify this interpretation one can calculate the scattered light intensity and see precisely how it scales with $n_L$ and $n_R$. This can be accomplished by using Eqs (\ref{Hr}) and (\ref{Vsr}) to obtain the Heisenberg equation of motion for $\hat{a}_{\bfk\lambda}$, 
\begin{eqnarray}
\label{dakdt}
	\ddt\hat{a}_{\bfk\lambda}&=&-i\omega_k\hat{a}_{\bfk\lambda}\nonumber\\
	&-&i\frac{\Omega}{\Delta}g^\ast_{\bfk\lambda}e^{-i\omega_Lt}\int d^3r\, e^{i(\bfk_L-\bfk)\cdot\bfr}\hPsi^\dag(\bfr)\hPsi(\bfr),
\end{eqnarray}
where we have substituted the adiabatic solution $\hpsi_e(\bfr)=\frac{\Omega}{\Delta}\hPsi(\bfr)$ . Substituting the mode expansion (\ref{elasticmodes}) and noting that $\hat{n}_L$ and $\hat{n}_R$ are constants of motion in the absence of tunneling gives upon time integration
\begin{eqnarray}
\label{akt}
	& &\hat{a}_{\bfk\lambda}(t)=\hat{a}_{\bfk\lambda}(0)e^{-i\omega_kt}
	-i\frac{\Omega}{\Delta}g^\ast_{\bfk\lambda} e^{-i(\omega_k+\omega_L)t/2}\nonumber\\
	& &\times\frac{\sin[(\omega_k-\omega_L)t/2]}{(\omega_k-\omega_L)/2}
	\left[e^{i(\bfk-\bfk_L)\cdot\frac{\bf{s}}{2}}\hat{n}_L+e^{-i(\bfk-\bfk_L)\cdot\frac{\bf{s}}{2}}\hat{n}_R\right].\nonumber\\
\end{eqnarray}
From this expression we can compute the mean number of scattered photons via $n_p(t)=\sum_{\bfk\lambda}\langle\hat{a}^\dag_{\bfk\lambda}\hat{a}_{\bfk\lambda}\rangle$, yielding
\begin{equation}
\label{npt}
	n_p(t)=\frac{|\Omega|^2}{\Delta^2}\Gamma\left[n_L^2+n_R^2+2\cos(\bfk_L\cdot{\bf s})\mu(\mathbf{s})\, n_Ln_R\right]\, t
\end{equation}
where we have converted the sum to an integral in the limit of infinite quantization volume and made the approximation $\sin^2(x t)/x^2\approx\pi t\, \delta(x)$ when integrating over $k$. This expression reveals an additional component to the scattering rate proportional to $\cos(\bfk_L\cdot{\bf s})\mu(\mathbf{s}) n_Ln_R$. This term is only significant when $\mu(\mathbf{s})\sim 1$, which requires that the mode separation be comparable to or less than the optical wavelength. In addition it depends on the exact spacing and orientation of the modes with respect to the polarization of the lasers, via the implied dependence of $\mu(\mathbf{s})$ on $\bf{s}$. We interpret this additional scattering as superradiance, as it arises from the imaginary part of the two-body dipole-dipole interaction. We see that in the limit as $s/\lambda_L\to0$ we have $\cos(\bfk_L\cdot{\bf s})\mu(\mathbf{s})\to1$ so that the total scattering rate is proportional to $N^2$. As stated previously, this means that the radiation field never acquires sufficient information to distinguish the states $|n\rangle$ and $|m\rangle$, thereforet no decoherence occurs.

\subsection{Dynamics under decoherence in elastic regime}
 In the previous section, we derived the master equation of the two-mode system in the presence of laser-induced atomic polarization. In this section we use numerical simulations to observe how the decoherence will affect the tunneling dynamics of our system and determine what level of decoherence 
can be tolerated during the dynamical generation of a cat state. In this section we will only focus on the decoherent dynamics in this elastic regime where the total atom number can be treated as a constant of motion.  In addition, we will assume that the separation of condensates are much larger than the laser wavelength, so that the dynamics  is determined Eq. (\ref{meqtd}) but with $\mu_\mathbf{s}=0$. Under those assumptions, it is straightforward to guess the coherence time $t_{coh}$ as
\begin{equation}
	t_{coh}=\frac{4 \Delta^2}{N^2|\Omega|^2\Gamma},
\label{tcoh}
\end{equation}  
by observing that thepeaks of the dynamically created cat-state are separated by $n_0=\pm N/4$. This is the time-scale on which coherence between states separated by $N/4$ should decay based on the decoherence rate (\ref{gammanm2}).
During the generation of the cat state, it is expected that if the coherence time $t_{coh}$ is large compared to $t_{cat}$, the cat state can still be produced. Otherwise, the system should evolve into a mixed-state rather than a pure cat-state.

\begin{figure}
\epsfig{figure=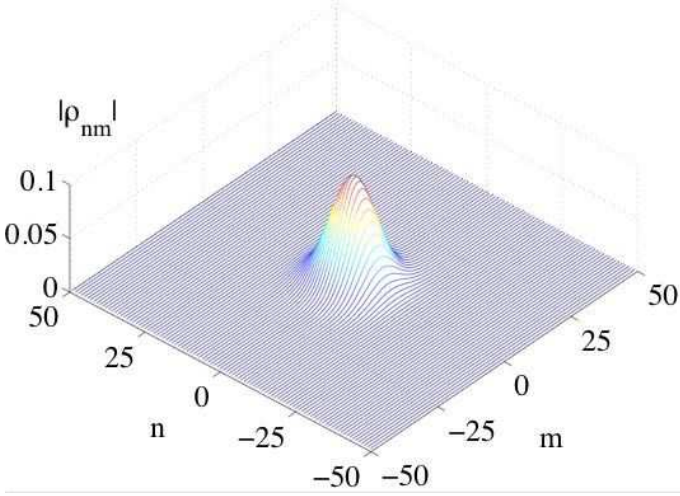, width=9.0cm}
\caption{Magnitude of the density matrix elements for the initial coherent state.\label{initialdensity}}
\end{figure}

\begin{figure}
\epsfig{figure=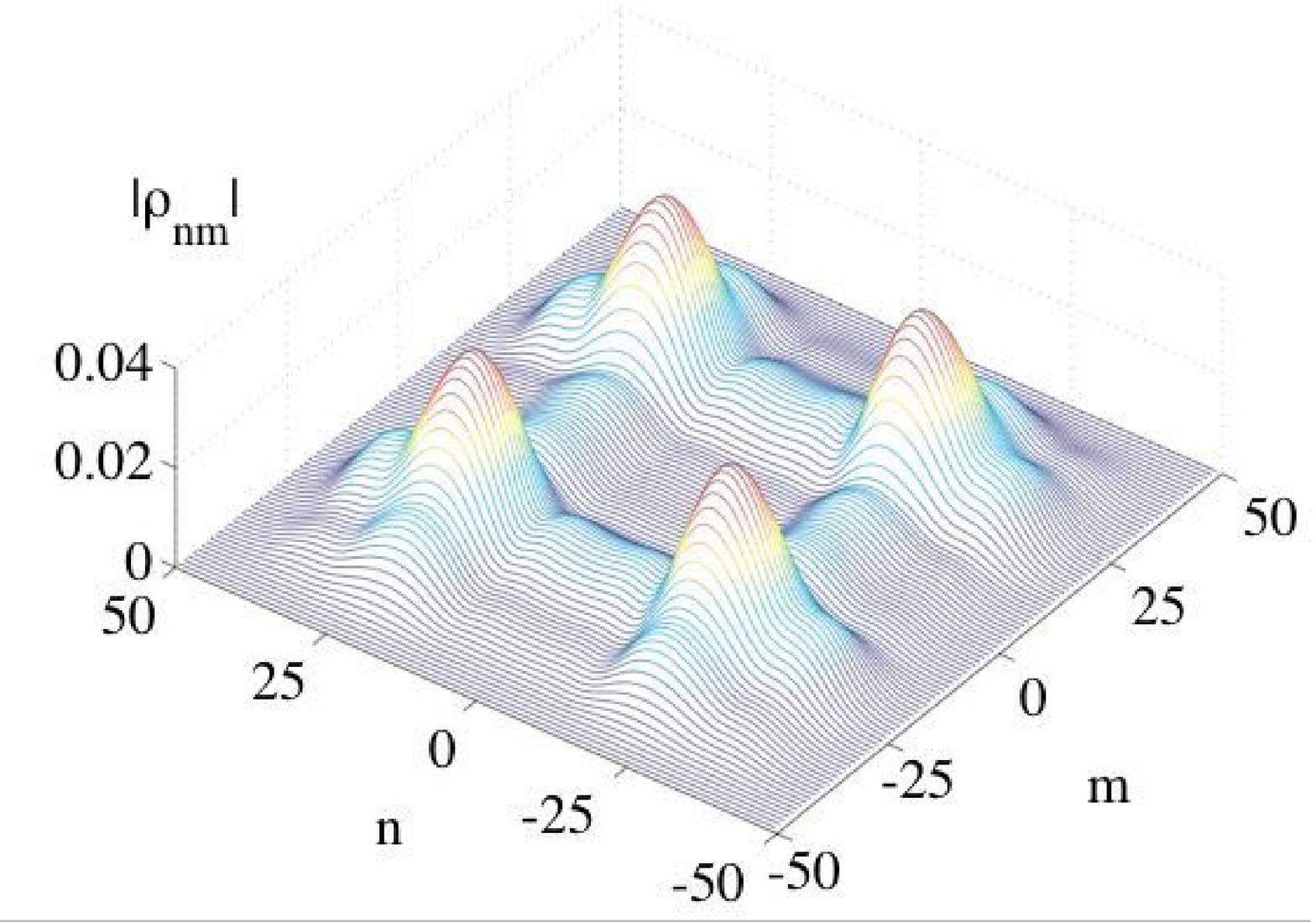, width=9.0cm}
\caption{Magnitude of the density matrix elements of the dynamical created cat state at $t=t_{cat}$ with no decoherence.\label{catdensity}}
\end{figure}

\begin{figure}
\epsfig{figure=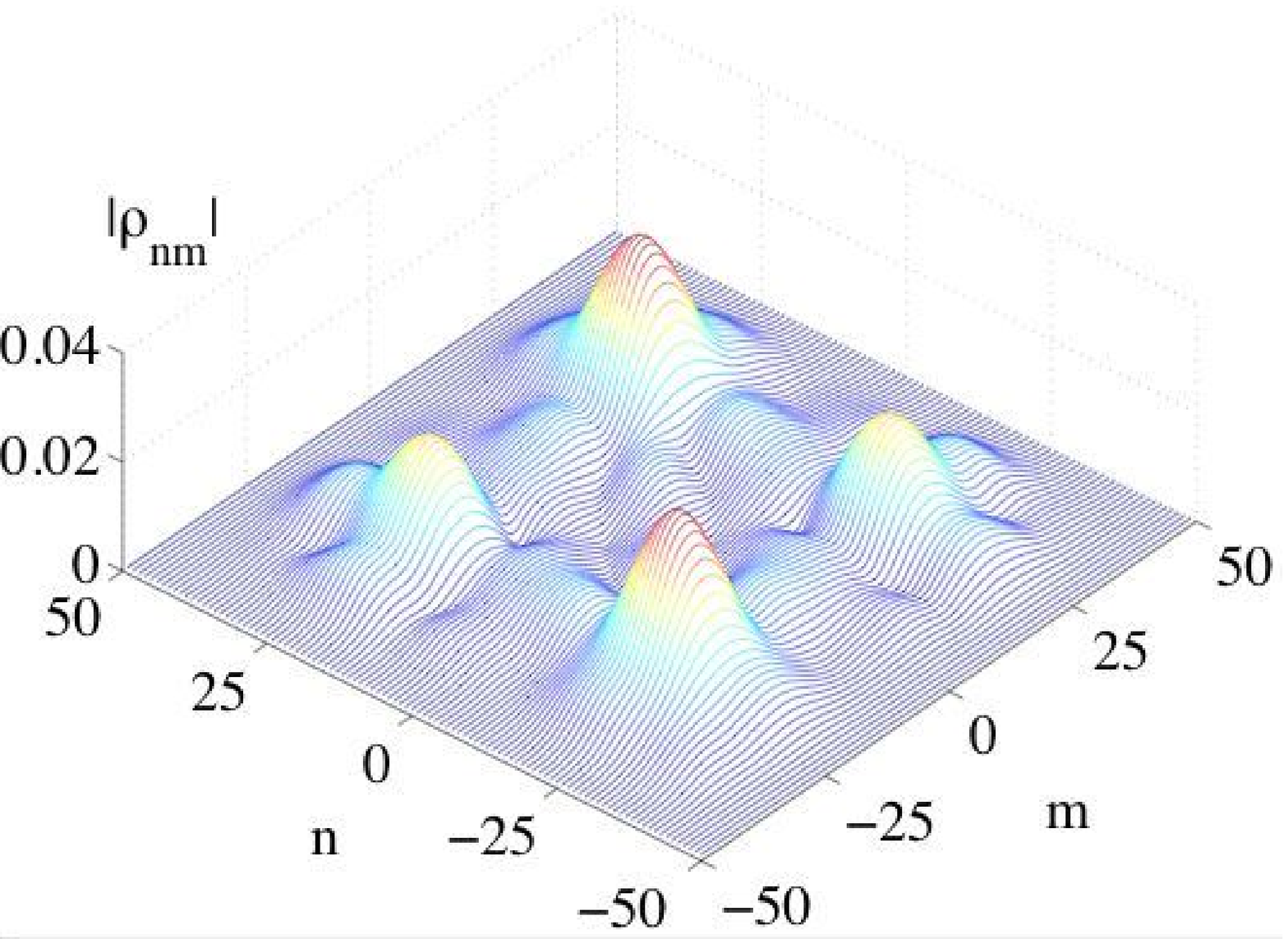, width=9.0cm}
\hbox{\hspace{0.8cm} (a)}
\epsfig{figure=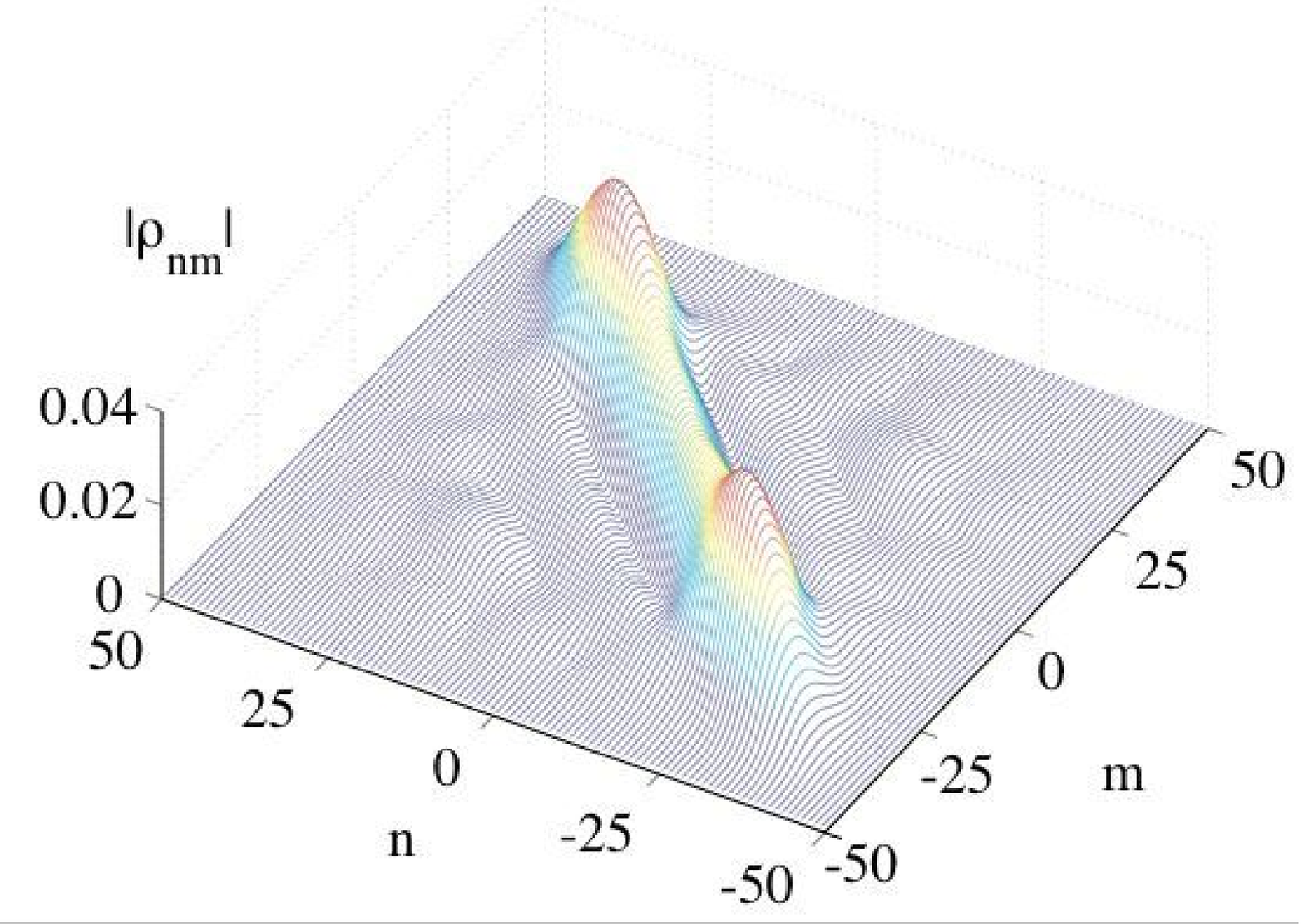, width=9.0cm}
\hbox{\hspace{0.8cm} (b)}
\caption{Magnitude of density matrix elements of the resulting state under weak decoherence (a) and strong decoherence (b). System parameters are chosen the same as in Fig.(\ref{formation-cat}), and the resulting states (a) (b) are measured at a time of $t_{cat}=14.5/\tau$.\label{dec-form}}
\end{figure}

The effects of decoherence on the dynamical evolution of cat-states is shown in Figures \ref{initialdensity}-\ref{dec-form}.
In Fig. \ref{initialdensity} we show the magnitude of the density-matrix elements for the initial coherent state at $t=0$. Under the condition of no decoherence, this state evolves into a cat-state at $t=t_{cat}$, which is shown in Fig. \ref{catdensity}. In Figures \ref{dec-form} (a) and (b) we show the resulting state for the cases of weak and strong decoherence, respectively. In Fig. \ref{dec-form}(a) the decoherence strength is chosen such that $t_{coh}=5t_{cat}$, while for the strong decoherence of Fig. \ref{dec-form} (b), we have $t_{coh}=t_{cat}$. It is seen that with $t_{coh}=5t_{cat}$ we can still create cat-like states, while as the decoherence strength increases up to $t_{coh}=t_{cat}$, the outcome state is a mixted state with no coherence between macroscopically distinguishable distributiions. Therefore, to successfully produce cat state, the coherence time  $t_{coh}$ should be much longer than the time need to produce cat state $t_{cat}$.

\begin{figure}
\epsfig{figure=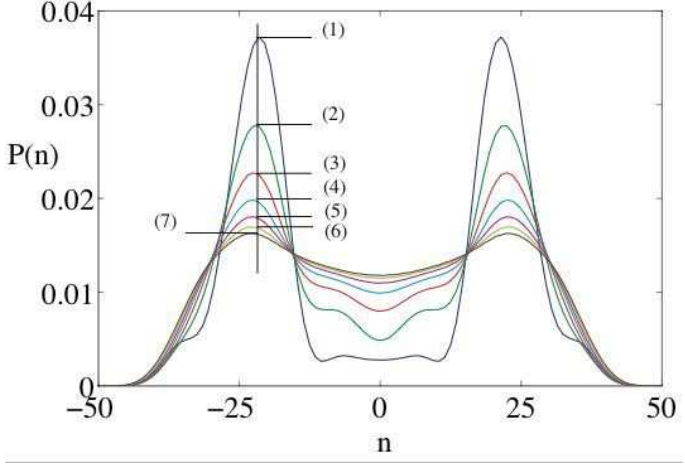, width=9.0cm}
\caption{Probability distribution $P(n)$ of the resulting states corresponding to different coherence times. Parameters are choose as in Fig. (\ref{dec-form}), while from line (1) to (7), the coherent time $t_{coh}$ is chosen as $\infty, 3 t_{cat}, 3 t_{cat}/2, t_{cat}, 3 t_{cat}/4, 3 t_{cat}/5, t_{cat}/2$, respectively. \label{dec-form-pn}}
\end{figure}

It is also quite interesting to observe from Fig. \ref{dec-form} that not only the off diagonal, but also the diagonal of density matrix  of the final states are affected by decoherence effect. In Figure \ref{dec-form-pn}, we plot the probability distribution $P(n)\equiv\rho_{nn}$ of several resulting states corresponding to different coherence times. In the figure we see that the probability distribution of the mixed states formed under strong decoherence iis significantly different from that of the pure cat state. In other words, if decoherence dominates the system dynamics, it will not yield the same $P(n)$. Thus the probability distribution alone supplies a direct way to verify the existence of cat state at time $t_{cat}$, without the need to  detect revivals. We note that we have so far only established this result for the elastic regime. We suspect that the elastic regime will be much more difficult to achieve experimentally than the inelastic regime, due to the necessity of sub-wavelength confinement.

\begin{figure}
\epsfig{figure=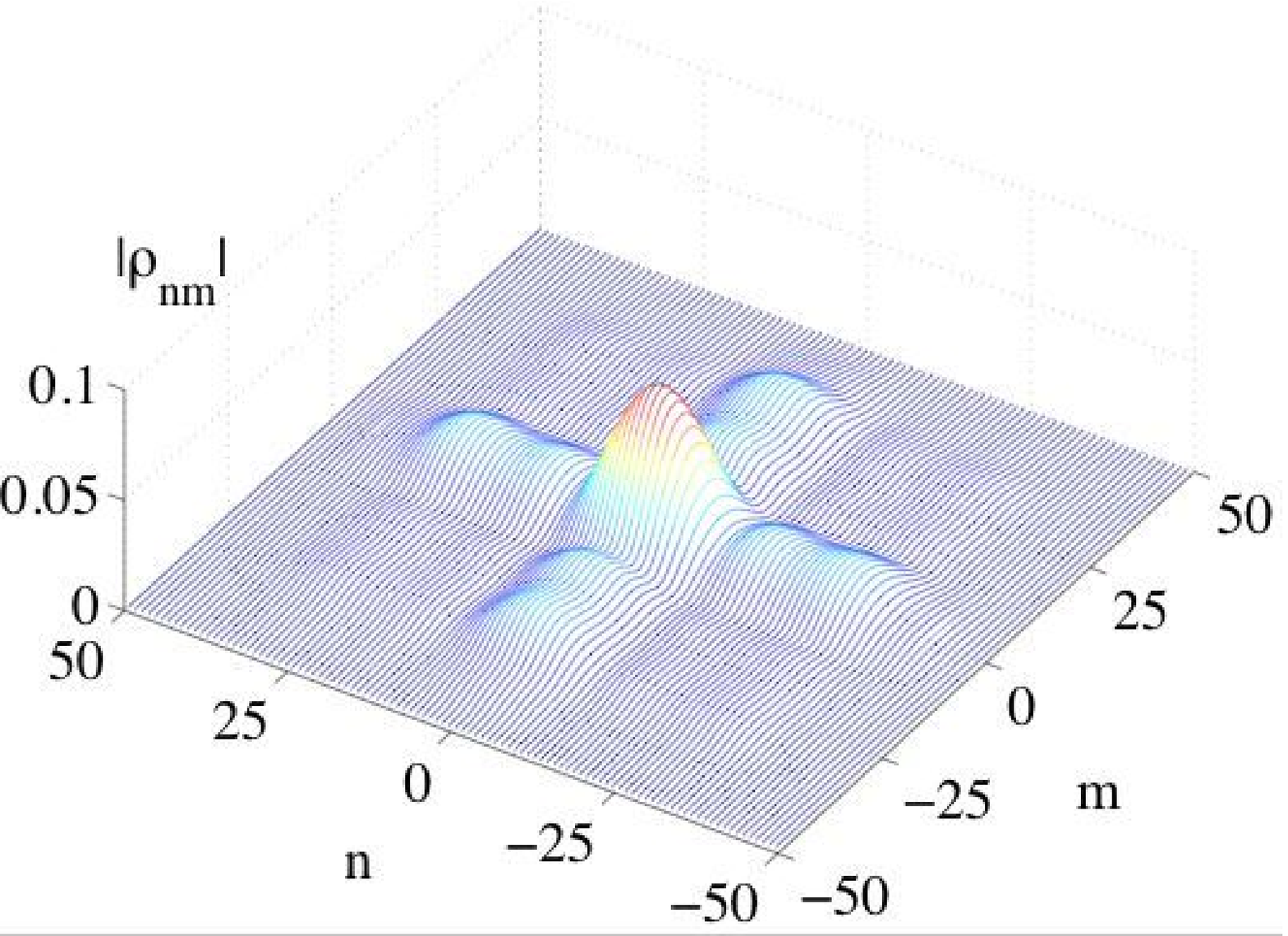, width=9.0cm}
\hbox{\hspace{0.8cm} (a)}
\epsfig{figure=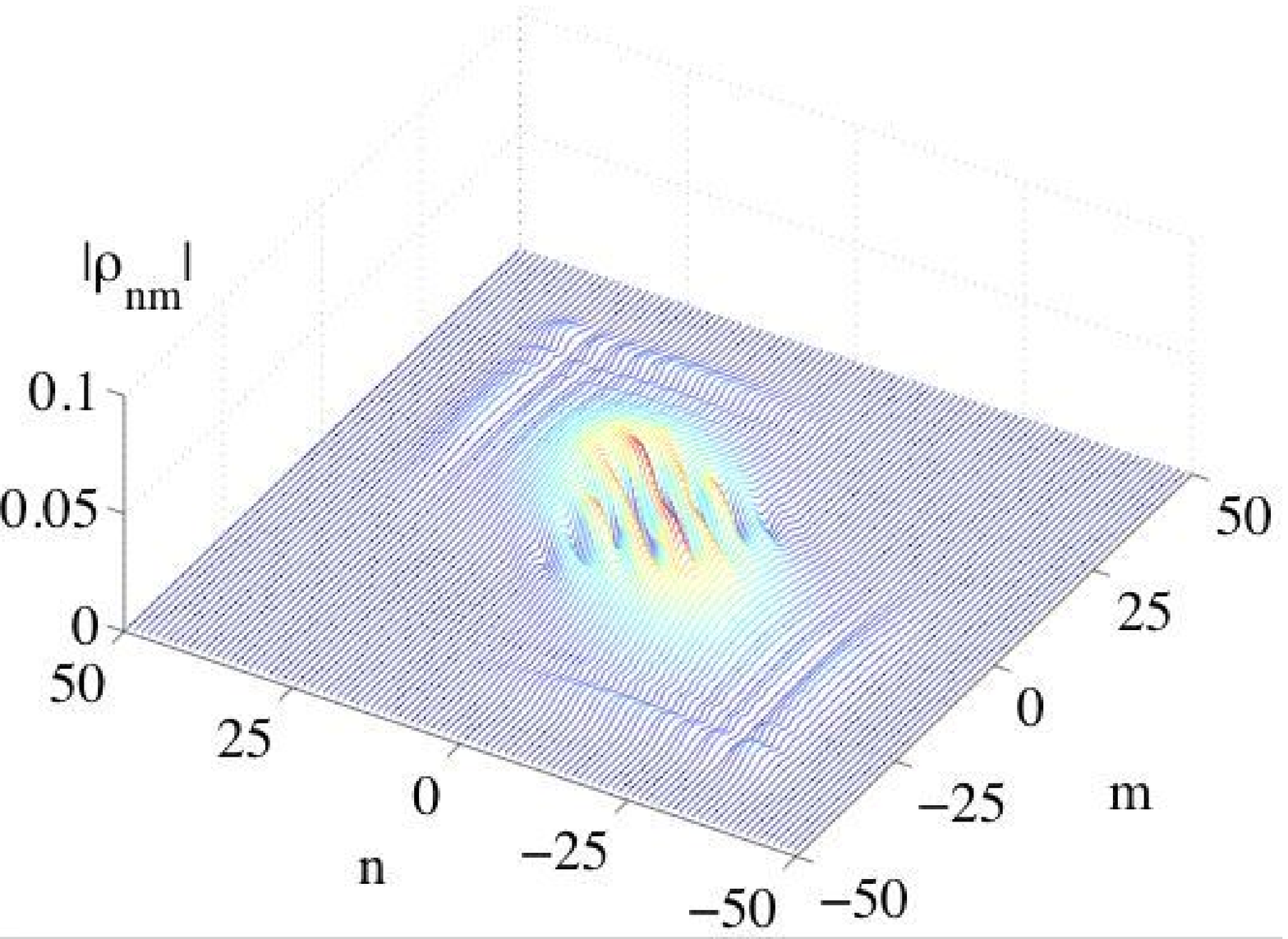, width=9.0cm}
\hbox{\hspace{0.8cm} (b)}
\epsfig{figure=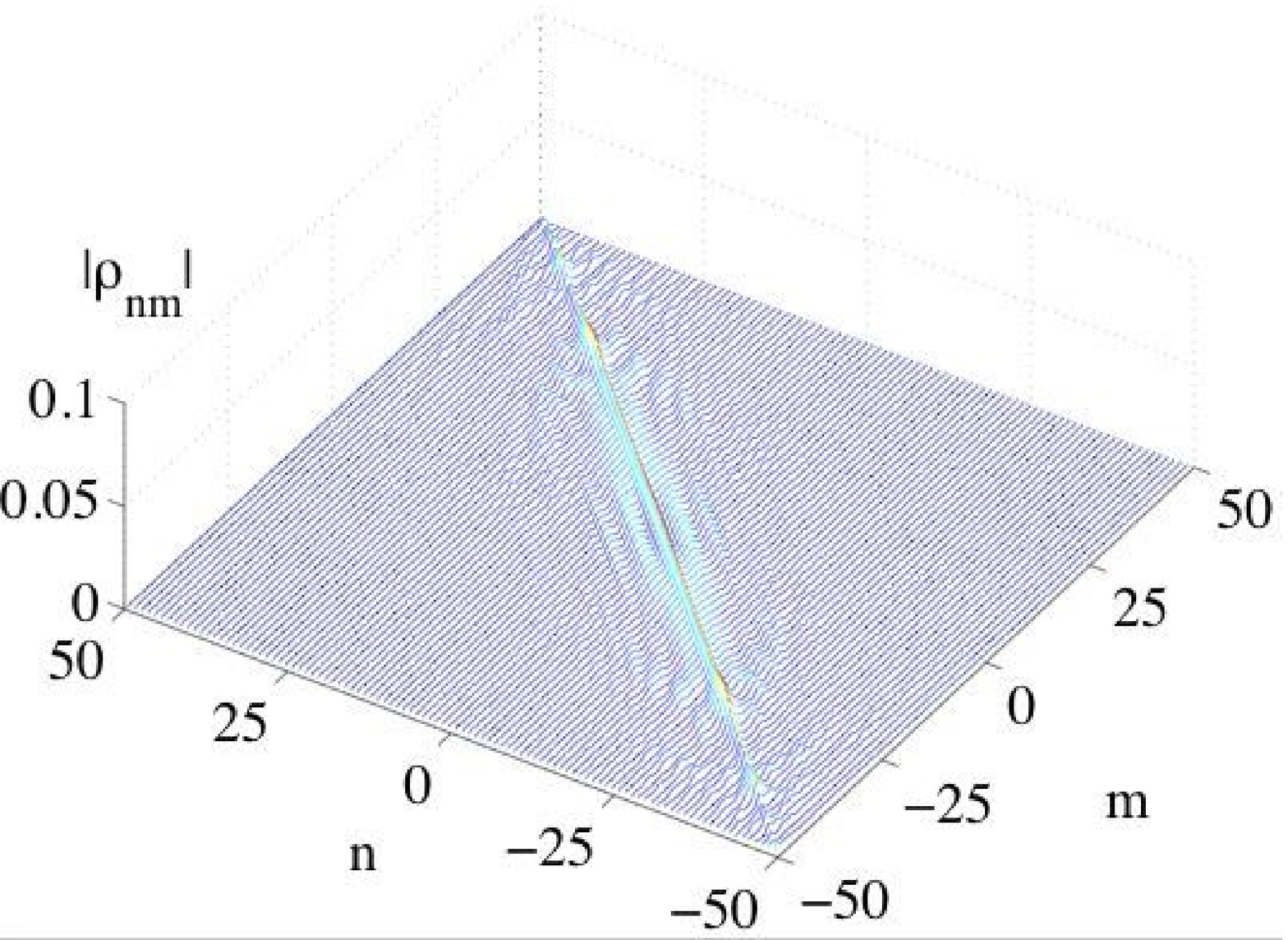, width=9.0cm}
\hbox{\hspace{0.8cm} (c)}
\caption{Effects of decoherence on the revival of the initial coherent state. Shown is the magnitude of the density matrix elements of the system at time $t=t_{cat}+t_{hold}+t_{rev}$ under no decoherence (a), the decoherent resulting state with $t_{hold}=t_{coh}$ (b),  and $t_{hold}=N^2 t_{coh}/4$ (c). System parameters are chosen the same as in Fig.(\ref{formation-cat}). \label{decoherence-revival}}
\end{figure}

In Figure \ref{decoherence-revival} we examine the effects of decoherence on the revival process, under the assumption that $t_{cat}<t_{coh}$, but assuming that the system is held at $\tau=0$ for a time $t_{hold}$, which may be longer than the coherence time. These figures show the magnitude of the density-matrix elements at time $t=t_{cat}+t_{hold}+t_{rev}$, calculated by direct numerical solution of the master equation (\ref{two-mode-master-delta}). In Figure \ref{decoherence-revival} (a) we show the magnitude of the density-matrix elements for the case $t_{hold}=0$ and for $t_{coh}=\infty$, i.e. no decoherence is present. Note that this figure corresponds to $t=21$ in Fig. \ref{revival-cat}. In Figure \ref{decoherence-revival} (b) we plot the magnitude of the resulting density matrix for the case $t_{hold}=t_{coh}=10t_{cat}$. This choice gives sufficient time to destroy coherence between fock states where $n$ differs by $N/2$, corresponding to the distance between peaks in the initial cat state. This shows that the revival is effectively destroyed by decoherence on the the timescale governed by the coherence time (\ref{tcoh}). This figure corresponds to the `partially incoherent state'  shown in Fig. \ref{partial-decoherence} at time $t=21$, in the sense that  the choice $t_{hold}=t_{coh}$ implies the condition $\sigma<I_{coh}<n_0$.  Lastly, Figure \ref{decoherence-revival} (c) shows the case $t_{hold}=\frac{N^2}{4}t_{coh}$ with $t_{coh}=10t_{cat}$ which gives sufficient time to destroy coherence between any fock states. We see this case that the system has collapsed onto an almost pure mixture of fock-states, and it is clear that no revival occurs. This figure corresponds to the totally incoherent state of Fig. \ref{complete-decoherence}.

\subsection{Loss and Decoherence in the inelastic regime}
\label{inelastic}

In this subsection we turn to the inelastic regime, which takes into account the finite size of the atomic modes, so that the scattering of laser photons is not entirely elastic due to photon recoil effects. The resulting losses from the two initially occupied modes should necessarily be accompanied by decoherence, which may or may not be sufficient to 'collapse'  the two-mode cat state. We focus solely on the case where the mode separation is large compared to the wavelength, so that we can safely ignore the cross terms in Eq. (\ref{two-mode-master-delta}). The master equation in this situation is found by substituting the two-mode expansion $\hPsi(\bfr)=\phi_L(\bfr)\hat{c}_L+\phi_R(\bfr)\hat{c}_R$, yielding
\begin{eqnarray}
\label{bimodal-master}
	\dot{\rho}_s(t) &=& -\frac{|\Omega|^2\Gamma}{2\Delta^2} \big[N\rho_s(t)+
        \xi[n^2_L+n_R^2-N]\rho_s(t) \nonumber\\
       &+& \xi[n_L\rho_s(t)n_L+n_R\rho_s(t) n_R]\big]+H.c.,
\end{eqnarray}
with $\xi$ defined in Eq. (\ref{xi}).  Deriving the equation of motion for the density matrix elements is then straightforward, giving
\begin{equation}
	\dot{\rho}_{nm}(t)=-\Gamma\frac{|\Omega|^2}{\Delta^2}[(1-\xi) N +\xi (n-m)^2]\rho_{nm}(t).
\end{equation}
Under the condition of spatially broad (relative to $\lambda_L$) condensates, $\xi\approx 0$, as seen if Figure \ref{fig-inelastic}, and this equation is reduced to
\begin{equation}
	\dot{\rho}_{nm}(t)=-\Gamma\frac{|\Omega|^2}{\Delta^2} N\rho_{nm}(t),
\end{equation} 
which shows that the decoherence rate is exactly equal to the atomic loss rate (\ref{dndtsimple}). The reason for this is that the loss of an atom takes us from the manifold with $N$ atoms to that with $N-1$ atoms. This is equivalent to $n\to n\pm1$ where the choice of $\pm$ depends on whether the atom was lost from the L or R mode (- for L and + for R). Thus one might suspect that a careful treatment of the scattering of the atom into the quasi-continuum of recoil modes would reveal that the loss of coherence on the matrix element $\rho_{nm}$ could be accompanied by a corresponding appearance of coherence on the matrix element $\rho_{n\pm1,m\pm 1}$ which lies in the $N-1$ manifold.

A complete quantum  description of this inelastic process should include both the initial cat-state modes as well as a quasi-continuum of recoil modes.We can therefore describe the initial unscattered systems with a density of product $\rho^N_{sr}=\rho^{N}\otimes\rho_r$, where $\rho^m_{sr}$ is the full density operator having $m$ atoms in the initial modes and $N-m$ atoms in the quasi-continuum, and  $\rho^m$ and $\rho_r^{N-m}$ are the corresponding reduced density matrices for the system and reservoir. After one photon is scattered, the full density matrix goes from $\rho^N_{sr}$ to $\rho^{N-1}_{sr}$, due to the transfer of one atom from the initial modes into the quasi-continuum. As the modes are well-separated, the scattered photon carries sufficient phase information into the environment to determine whether the atom recoils from the L or the R mode. Assuming that atoms from L and R modes scatter into different manifolds of recoil modes, the reduced density operator of the two-mode system, obtained by tracing out the quasi-continuum modes, becomes
\begin{equation}
\label{collapse}
	\rho^{N-1}=\frac{1}{\mathcal{N}}(c_L\rho^N c^\dagger_L+c_R\rho^N c^\dagger_R)
\end{equation}
 with $\mathcal{N}$ the normalization factor. In other words the system collapses onto an incoherence mixture of the state with one atom lost from mode L and the state with one atom lost from mode R.
 
 As an example, we now consider a $N$ particle cat state of 
\begin{equation}
	\rho^N=|\Psi_c\rangle\langle \Psi_c|.
\end{equation}
and to distinguish cat states with different total atom numbers, here we use a new notation 
\begin{equation} 
\label{less-ex}
 	|\Psi_c\rangle=\frac{1}{\sqrt{2}}(|N/4,3N/4\rangle+|3N/4,N/4\rangle),
\end{equation}
 where $|n_L,n_R\rangle$ stands for a Fock basis with $n_L$ atoms in left mode and $n_R$ in the right.  This cat state is collapsed after one photon-detection into a statistical mixture of two $N-1$ particle states,
\begin{equation}
	\rho^{N-1}=\frac{1}{2}(|\Psi_1\rangle\langle \Psi_1|+|\Psi_2\rangle\langle \Psi_2|)
\end{equation}
where 
\begin{eqnarray}
	|\Psi_1\rangle=\frac{1}{\sqrt{2}}(|N/4-1,3N/4\rangle+|3N/4-1,N/4\rangle), \\
	|\Psi_2\rangle=\frac{1}{\sqrt{2}}(|N/4,3N/4-1\rangle+|3N/4,N/4-1\rangle).
\end{eqnarray}
The important point here is that each of these states is still a good cat-like state, so that while we don't know which one the system has collapsed into, we do know that the system remains in a good cat-like state. The means that the effect of scattering a photon and losing an atom may not have a significant detrimental effect on cat-like states. Similarly, after a second photon is scattered, the system will collapse into a statistical mixture of four $N-2$ particle cat states, and so on. We note, however, that for the 'perfect' cat state $\left[|N0\rangle+|0N\rangle\right]/\sqrt{2}$ a single scattering results in a mixture of the states $|(N-1)0\rangle$ and $|0(N-1)\rangle$, neither of which is itself a cat-state. This means that while the ideal cat state is so fragile that  a single scattered photon is sufficient to collapse that cat onto all-left or all-right states, the cat-like states we are considering should be significantly more robust, giving them a significant advantage provided they are still suitable for whatever application is desired.

\begin{figure}
\epsfig{figure=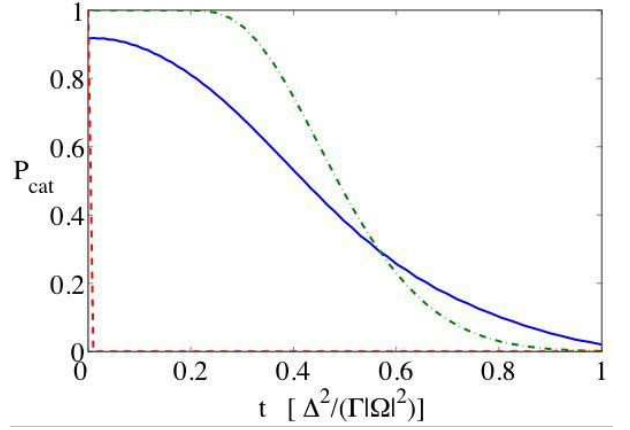, width=8.0cm}
 \caption{The evolution of the projector $P_{cat}$ under decoherence die to inelastic atom loss. The curves correspond to the dynamically-created cat state (Blue solid line), the perfect cat state (Red dashed line) and less extreme cat state (\ref{less-ex}) (Green dashed-dotted line). The figure plots the estimated probability to remain in a cat-like state versus time. Starting from 100 atoms, it is estimated that $63$ atoms are lost by the final time $t=\frac{\Delta^2}{|\Omega|^2\Gamma}$.    
\label{decproj}}.
\end{figure}

To show the  dynamics under decoherence induced by these inelastic condensate losses, in Fig. \ref{decproj}, we plot the time evolution of the expectation value of the projector $P_{cat}$, defined in (\ref{projection-cat}). Three initial states are shown, the extreme cat state $\left[|N,0\rangle+|0N\rangle\right]/\sqrt{2}$, the less extreme cat state, $\left[|\frac{N}{4}\frac{3N}{4}\rangle+|\frac{3N}{4},\frac{N}{4}\rangle\right]/\sqrt{2}$ and the wave-packet cat-states produced via dynamical evolution as in Section \ref{dynev}. 

Figure \ref{decproj}  is obtained by assuming a single photon is detected in each time interval $\Delta t=\frac{\Delta^2}{|\Omega|^2\Gamma N(t)}$, where $N(t)$ is the number of remaining atoms at time $t$. After each scattering a new density matrix is obtained by applying Eq. (\ref{collapse}). We note that the use of (\ref{collapse}) automatically weighs the probabilities that the scattering occurs from the L or R mode due to the action of the annihilation operators being proportional to the square root of the mode occupation. The resulting density matrix is then diagonalized and written
\begin{equation}
\rho=\sum_{i} ~P_i |\Psi_i\rangle \langle\Psi_i|,
\end{equation}
where $|\Psi_i\rangle$ denotes the $i$th eigenmode, with $P_i$ its statistical weights over the mixture state. The resulting eigenmodes $|\Psi_i\rangle$'s are then analyzed to determine whether or not they are sufficiently cat-like. This is accomplished by first assigning a value of zero to any state where $\sum_{n>0}|c_n|^2\neq\sum_{n<0}|c_n|^2$, where $c_n$ is again the probability amplitudes for the state $|n\rangle$. The states which survive this test are then assigned a value equal to the expectation value of the projector $P_{cat}$. The assigned values for each eigenmode of the density matrix at time t are then averaged together  and the results are plotted versus time. 
\begin{figure}
\epsfig{figure=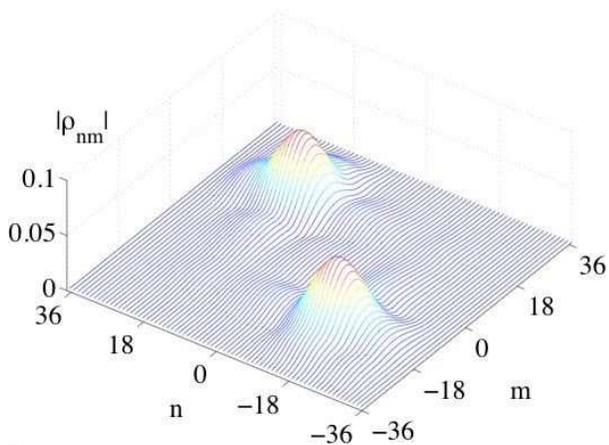, width=8.0cm}
 \caption{Magnitude of the resulting density matrix elements corresponding to Fig. \ref{decproj} at time $t=0.3 \frac{\Delta^2}{|\Omega|^2\Gamma}$.      
\label{decrho}}.
\end{figure}
\begin{figure}
\epsfig{figure=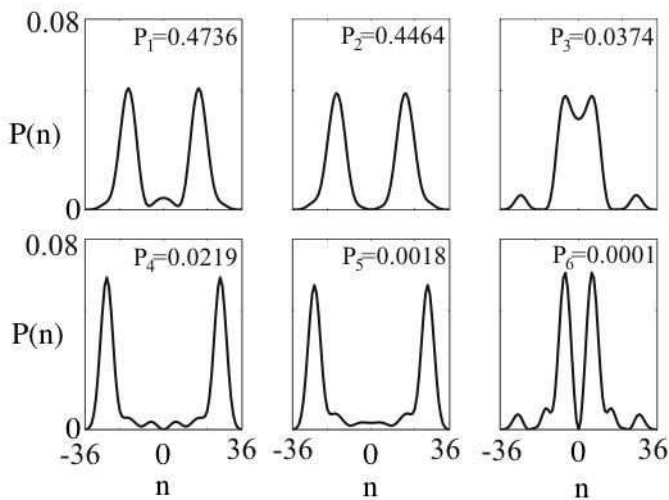, width=9.0cm}
 \caption{Probability distributions $P(n)$ of the six leading eigenmodes of the density matrix shown in Fig. (\ref{decrho}). 
 \label{6modes}     
\label{decwf}}.
\end{figure}

From the figure, we verify that the perfect cat state is completely collapsed immediately after one photon scattering. In contrast,  the less extreme cat states has a high probability to remain in a cat-like state even after many scatterings, and, the dynamically created cat-states are similarly robust against decoherence. We note that in the figure out of an initial population of $N=100$ atoms, $N=37$ remain at the end-time $t= \frac{\Delta^2}{|\Omega|^2\Gamma }$.

In Fig.  \ref{decrho}, we plot the magnitude of the  density matrix after a time of $t_{hold}=0.3 \frac{\Delta^2}{|\Omega|^2\Gamma}$ as in Fig. (\ref{decproj}) (Blue solid line). In this figure, an initial cat state is formed at time $t_{cat}$ via dynamic evolution with no decoherence. The state is then held with $g=0$ and $\tau=0$ for time $t_{hold}$ with photon-scattering acting as a gradual source of decoherence. The figure appears to show a collapse of the reduced density matrix onto a statistical mixture of left-centered and right-centered peaks. However, diagonalization of this matrix reveals that this is not the case. Based on numerical diagonalization, we plot the major eigenmodes $|\Psi_i\rangle$ of this density matrix in Fig. \ref{6modes}. From the picture, some of these resulting eigenmodes, such as $|\Psi_i\rangle$'s with $i=1,2,4,5$ are still cat-like, in a sense that separation between two peaks are larger than the width of either peak. While for some others, like $|\Psi_3\rangle$ and $|\Psi_6\rangle$, the separation between peaks are not larger than the width of each peak, and thus they are not cat-like. As more atoms are scattered, the number of remaining atoms decreases, and hence the
maximum separation of cat-like states will become will necessarily decrease. Because we have chosen projector $\hat{P}_{cat}$ to be fixed-width filter, those cat-like state with small separations between peaks may not be picked by $\hat{P}_{cat}$. As a result, $P_{cat}$ decreases to almost zero before all atoms are lost, as seen in Fig. (\ref{decproj}). Lastly, we note that In this section we did not show the effect of inelastic decoherence on the tunneling dynamics during the formation of the cat state. This can be done by the standard quantum jump method in Mento-Carlo simulation, which we plan to carry out in future work. 

\section{Dephasing induced by atomic collision}
\label{dephasing}
In this section we consider the effects of atom-atom collisions on a cat-state which is 'frozen' by reducing the tunneling strength, $\tau$, to zero. If during this hold state the atomic interaction strength, $g$, is nonzero, the system will undergo a dephasing process \cite {A. Imamoglu1997-1,Greiner2002},  whose dynamics is determined by Hamiltonian (\ref{no-tunneling}). The main effect of this dephasing process is that, while it will not collapse the coherence of the cat-state, once the state is dephased, it cannot evolve back to the initial coherent state. In this way, a detection scheme based on revivals will incorrectly create the appearance of collape. Therefore, in general, the dephasing effect needs to be suppressed by making the atomic interaction strength $g$, and/or the hold time small. 
\begin{figure}
\epsfig{figure=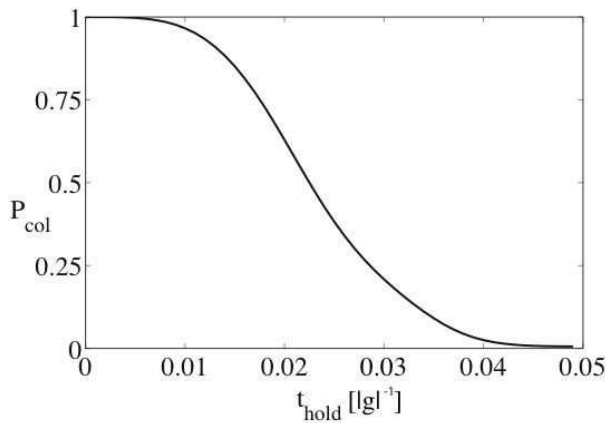, width=8.0cm}
 \caption{Dephasing process of cat state with nonzero atomic interaction. The collapse is measured by $P_{col}=tr\{\rho_c\rho(t)\}$, where $\rho_c$ and $\rho(t)$ are the densities of the initial cat state and dephased state after a hold time $t_{hold}$ respectively. Parameters are given the same as in Fig. \ref{formation-cat} while $t_{hold}$ has units of $1/|g|$.  \label{dephase}}.
\end{figure}
  
In studying the dephasing effect, first, we note that at each period of $\pi/g$, the relative phase of each $c_n$ is exactly restored and thus the many-body states is revived.  While between two consequent revivals, nonzero relative phases are introduced to Fock basis, and the cat state will collapse. In Fig. \ref{dephase}, we show such a rapid dephasing process, where the cat-state decays with a collapse time around $t_{col}= 0.02/|g|$, determined from numerical simulations.

To obtain an analytic estimate of the collapse time scale, we assume that our system is prepared in a Gaussian-like cat state (\ref{cat-Gaussian}), which is a good approximation of the realistic cat state in our system. And then, we associate the collapse and revival of our bimodal system with the evolution of one-body two-mode correlation function $\Lambda$,
\begin{equation}
\label{correlated}
	\Lambda=\langle \texttt{cat}| \hat{a}^\dagger_L\hat{a}_R | \texttt{cat}\rangle. 
\end{equation}
At short time scale, the correlation function (\ref{correlated}) is obtained, under the condition of that the Gaussians are well peaked around $n=\pm n_0$, to a good approximation \cite{A. Imamoglu1997-1}, 
\begin{equation}
	\Lambda(t)=\Lambda(0) e^{-(g\sigma t)^2}
 \end{equation}
And therefore, the collapse time $t_{col}$ can be estimated by
\begin{equation}\label{collapse-time}
	t_{col}=\frac{1}{2g\sigma}. 
\end{equation}

\begin{figure}
\epsfig{figure=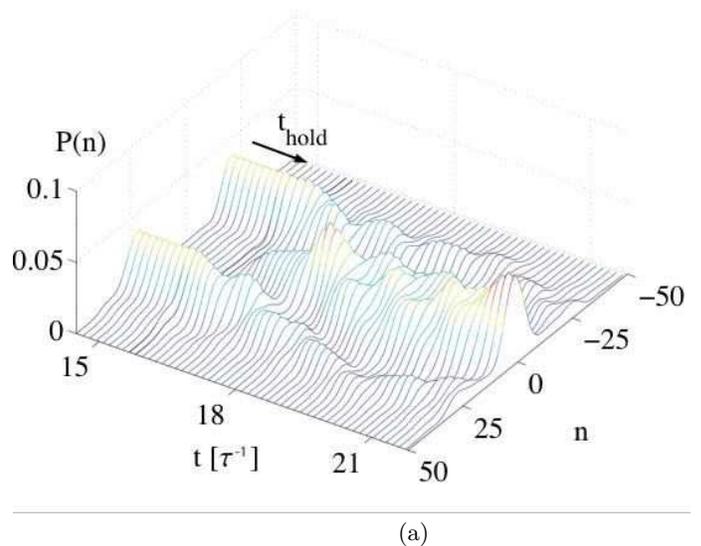, width=9.0cm}
\hbox{\hspace{2.0cm}(a)}
\epsfig{figure=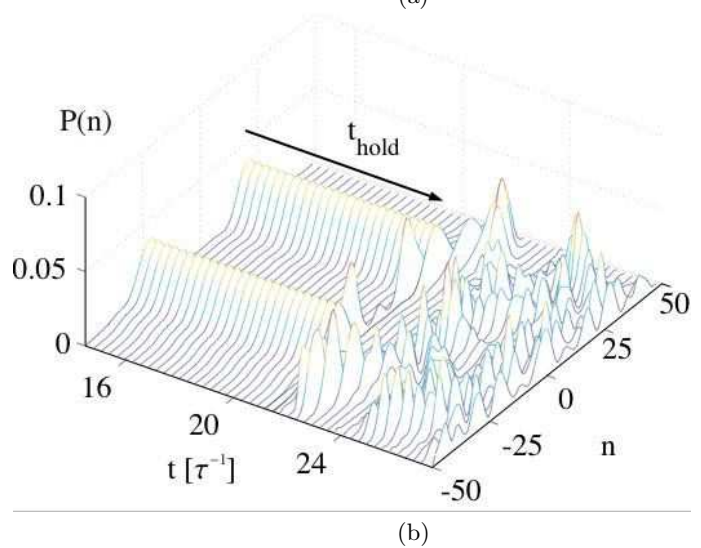, width=9.0cm}
 \hbox{\hspace{2.0cm}(b)}
 \caption{Dynamic evolution of dephased cat state after tuning on the tunneling. The parameters are chosen as in Fig. \ref{formation-cat} with a collapse time of $t_{col}=1.5/\tau$. In the first picture (a), the hold time is set to be $t_{hold}=t_{col}$, while for (b), we set $t_{hold}=5t_{col}$. It is shown that for (a), the system still revives to the initial state, while in (b), the initial state can not be restored.
  \label{dephase-revival}}.
\end{figure}

As discuss above, the collisional dephasing will affect the revival process. Generally speaking, if the holding time $t_{hold}$ is less than collapse time $t_{col}$, the system can still evolve back to the initial state as shown in Fig. \ref{dephase-revival} (a). But under the condition of $t_{hold}>t_{col}$, the cat state is dephased, and the coherent state can not be restored. Rather it will evolve to a disordered state as shown Fig. \ref{dephase-revival} (b).  Therefore, for the success of the detection scheme, one should make sure that the dephasing collapse time is longer than the hold time and the revival process is only affected by decoherence. 

\section{Conclusion}
\label{conclusion}

We have  considered the double-well BEC system and proposed that cat-like states can be formed by switching the sign of the interaction strength from positive to negative using a Feshbach resonance. We examined the possibility of either adiabatic or sudden switching with the following results. In the adiabatic case, the ground state evolves into a macroscopic superposition state with all of the atoms collectively occupying either the left or right well. However, by examining the degeneracy of the ground state, we conclude that the cat state is unstable against small perturbations and may readily evolve into a localized state. For the case of a sudden change in the scattering length, we find that an initial condensate state evolves dynamically into a double-peaked superposition of number states, with one peak corresponding to the majority of atoms being in one well, and the other peak corresponding to the majority being in the opposite well. 

We have shown that this process is stable against perturbations in the initial state, as well as in the control parameters. For this reason we believe that the dynamical evolution scheme is more likely to be demonstrated experimentally. In addition, we have demonstrated that continuing the evolution after the cat-state formation results in  a nearly complete revival of the initial state. However, this revival does not occur if the cat-state has collapsed into an incoherent mixed state, so that this revival can be taken as proof of coherence of the initial cat-state. We note that the states formed via adiabatic switching are maximum `all or nothing' cat states,  while those formed dynamically are less distinct cat-like states corresponding to a double-peaked distribution in Fock-space.

The effects of decoherence due to spontaneous scattering of laser photons have also been studied in detail.
We have determined that decoherence is strongest for the case where the single-photon momentum recoil is not sufficient to remove the atom from its initial state, due to  Bose-stimulation as the atom remains in an strongly-occupied atomic-field mode. In the opposite regime, where the scattering rate is smaller by a factor of $N$, $N$ being the atom number, we find that the maximally entangled `all or nothing' cat states are destroyed by the scattering of a single photon, while the dynamically-formed less-extreme cat-like states can survive multiple scatterings without leaving the subspace of cat-like states, in analogy to the effect of atomic collision losses studied in \cite{Micheli2003}. Because of this enhanced survivability, the less-extreme cat-states may be more useful for applications, providing they are suitable for the desired protocol. 

In this context of `freezing' the cat states for a prolonged time, we have also considered the de-phasing effects of atom-atom interactions during this stage, and conclude that collisions can mimic the effects of decoherence with regard to a revival-type measurement, however with regard to applications, they may or may not be detrimental, depending on the type of interaction and/or measurements involved.

We note that in the present proposal, there are three different time scales which we have established: the time need to create cat state $t_{cat}$, the coherence time $t_{coh}$ and the collisional dephasing time $t_{col}$. First, to generate the cat state, the coherence time scale $t_{coh}$ must longer than $t_{cat}$. While later, in order to hold the cat state for later use in an application, a hold time $t_{hold}$ shorter than $t_{coh}$ is required to maintain the macroscopic superposition state. And finally, another condition $t_{col} > t_{hold}$ should be satisfied to avoid the dephasing quasi-collapse, if such dephasing is detrimental to the desired application. The coherence time, $t_{coh}$, is tunable by adjusting the laser intensity and detuning, while the dephasing time $t_{dephase}$ can be adjusted if the interaction strength $g$ is modified during the hold stage.

\section{Appendix}
\label{adelim}
To eliminate the excited state in the master equation (\ref{reduced-master}), we note that for a far-off-resonant laser field, the excitation rate in BEC is extremely low, which enables us to expand the density matrix to a good approximation as
\begin{equation}
	\rho \approx \rho_{00}+\rho_{01}+\rho_{10}+\rho_{11},
\end{equation}
where $ \rho_{00}$ and $\rho_{11}$  stand for the density matrix of none-excitation and one excitation state respectively, while $\rho_{01}$ and $\rho_{10}$ describe coherence between these two manifolds. In fact, it is convenient to define a projection operator $\hat{P}_j$, which is a projection into a subspace with $j$ atomic excitations of the BEC system, such that
\begin{equation}
	\rho_{ij}=\hat{P}_i \rho \hat{P}_j.
\end{equation}
With this expansion, one obtains
\begin{eqnarray}
\label{rho00}
 	\dot{\rho}_{00}&=&-i \int d^3 r \,\Omega(\mathbf{r})\hat{\Psi}^\dagger_g(\mathbf{r})\hat{\Psi}_e(\mathbf{r}) \rho_{01}+ \nonumber \\
	                   & &\int d^3 rd^3r^{'} \, L(\mathbf{r}-\mathbf{r}') 
	                  \hat{\Psi}^\dagger_g(\mathbf{r})\hat{\Psi}_e(\mathbf{r})\rho_{11}\hat{\Psi}^\dagger_e(\mathbf{r}')\hat{\Psi}_g(\mathbf{r}')
	                  \nonumber \\
	                  & &+H.c.,
\end{eqnarray}
\begin{eqnarray}
\label{rho10}
 	\dot{\rho}_{10}&=&-i \int d^3 r \,\Omega^{*}(\mathbf{r})\hat{\Psi}^\dagger_e(\mathbf{r})\hat{\Psi}_g(\mathbf{r}) \rho_{00} \nonumber \\
	                   & &  +i\int d^3 r \,\Omega^{*}(\mathbf{r})\rho_{11}\hat{\Psi}^\dagger_e(\mathbf{r})\hat{\Psi}_g(\mathbf{r})  \nonumber \\
                          & &-\int d^3 rd^3r^{'} \, L(\mathbf{r}-\mathbf{r}') 
	                  \hat{\Psi}^\dagger_e(\mathbf{r})\hat{\Psi}_g(\mathbf{r})\hat{\Psi}^\dagger_g(\mathbf{r}')\hat{\Psi}_e(\mathbf{r}')\rho_{10} \nonumber \\
	                   & & -i\Delta \rho_{10},
\end{eqnarray}
\begin{equation}
\label{rho01}
 	\dot{\rho}_{01} =\dot{\rho}^\dagger_{10},
\end{equation}
and
\begin{eqnarray}
\label{rho11}
 	\dot{\rho}_{10}&=&-i \int d^3 r \,\Omega^{*}(\mathbf{r})\hat{\Psi}^\dagger_e(\mathbf{r})\hat{\Psi}_g(\mathbf{r}) \rho_{01} \nonumber \\
	                     & &-\int d^3 rd^3r^{'} \, L(\mathbf{r}-\mathbf{r}') 
	                  \hat{\Psi}^\dagger_e(\mathbf{r})\hat{\Psi}_g(\mathbf{r})\hat{\Psi}^\dagger_g(\mathbf{r}')\hat{\Psi}_e(\mathbf{r}')\rho_{11}
	                  \nonumber \\
	                  & &+H.c.. 
 \end{eqnarray}
In the assumption of low excitation rate, one can adiabatically eliminate $\rho_{01}$ and $\rho_{10}$ by demanding $\dot{\rho}_{01} \approx 0$ and $\dot{\rho}_{10} \approx 0$ respectively, yielding
\begin{eqnarray}
\label{rho10-solution}
 	\rho_{10}&=& -\int d^3 r \,\frac{\Omega^{*}(\mathbf{r})}{\Delta}
	                   \hat{\Psi}^\dagger_e(\mathbf{r})\hat{\Psi}_g(\mathbf{r}) \rho_{00}+ \nonumber \\
	                   & & +\int d^3 r \,\frac{\Omega^{*}(\mathbf{r})}{\Delta}
	                  \rho_{11} \hat{\Psi}^\dagger_e(\mathbf{r})\hat{\Psi}_g(\mathbf{r}) \nonumber \\
	                    & & i\int d^3 rd^3r^{'} \, \frac{L(\mathbf{r}-\mathbf{r}')}{\Delta} 
	                  \hat{\Psi}^\dagger_e(\mathbf{r})\hat{\Psi}_g(\mathbf{r})\hat{\Psi}^\dagger_g(\mathbf{r}')\hat{\Psi}_e(\mathbf{r}')\rho_{10}
	                  \nonumber \\
	\end{eqnarray}
Solving this equation perturbatively up to the order $\frac{1}{\Delta^2}$, one obtains
\begin{widetext}
\begin{eqnarray}
\label{rho10-solution-perturbation}
 	\rho_{10}&=& -\int d^3 r \,\frac{\Omega^{*}(\mathbf{r})}{\Delta}
	                   \hat{\Psi}^\dagger_e(\mathbf{r})\hat{\Psi}_g(\mathbf{r}) \rho_{00}+
	                   +\int d^3 r \,\frac{\Omega^{*}(\mathbf{r})}{\Delta}
	                  \rho_{11} \hat{\Psi}^\dagger_e(\mathbf{r})\hat{\Psi}_g(\mathbf{r}) \nonumber \\
	                    & & -i\int d^3 rd^3r^{'}d^3r^{''} \, \frac{L(\mathbf{r}-\mathbf{r}')\Omega^{*}(\mathbf{r^{''}})}{\Delta^2} 
	                  \hat{\Psi}^\dagger_e(\mathbf{r})\hat{\Psi}_g(\mathbf{r})\hat{\Psi}^\dagger_g(\mathbf{r}')\hat{\Psi}_e(\mathbf{r}')
	                   \hat{\Psi}^\dagger_e(\mathbf{r}^{''})\hat{\Psi}_g(\mathbf{r}^{''})\rho_{00}	                   
	                  \nonumber \\
	                   & & +i\int d^3 rd^3r^{'}d^3r^{''} \, \frac{L(\mathbf{r}-\mathbf{r}')\Omega^{*}(\mathbf{r^{''}})}{\Delta^2} 
	                  \hat{\Psi}^\dagger_e(\mathbf{r})\hat{\Psi}_g(\mathbf{r})\hat{\Psi}^\dagger_g(\mathbf{r}')\hat{\Psi}_e(\mathbf{r}')\rho_{11}
	                   \hat{\Psi}^\dagger_e(\mathbf{r}^{''})\hat{\Psi}_g(\mathbf{r}^{''}),
\end{eqnarray}
\end{widetext} 	
and $\rho_{01}=\rho_{01}^\dagger$.
Inserting this result for $\rho_{01}, \rho_{10}$  back into Eq. (\ref{rho11}), we find
\begin{widetext}
\begin{eqnarray}
\label{rho11-new}
 	\dot{\rho}_{11}&=& i\int d^3 r d^3 r^{'} \,\frac{\Omega^{*}(\mathbf{r})\Omega(\mathbf{r}')}{\Delta}
	                   \hat{\Psi}^\dagger_e(\mathbf{r})\hat{\Psi}_g(\mathbf{r}) \rho_{00}  \hat{\Psi}^\dagger_g(\mathbf{r}')\hat{\Psi}_e(\mathbf{r}')
	                   -i\int d^3 r d^3 r^{'} \,\frac{\Omega^{*}(\mathbf{r})\Omega(\mathbf{r}')}{\Delta}
	                   \hat{\Psi}^\dagger_e(\mathbf{r})\hat{\Psi}_g(\mathbf{r}) \hat{\Psi}^\dagger_g(\mathbf{r}')\hat{\Psi}_e(\mathbf{r}')\rho_{11} 
	                  \nonumber \\
	     	             & & +\int d^3 rd^3r^{'}d^3r^{''}d^3r^{'''} \, \frac{L^{*}(\mathbf{r}''-\mathbf{r}')\Omega^{*}(\mathbf{r})
		             \Omega(\mathbf{r}^{'''})}{\Delta^2} 
	                  \hat{\Psi}^\dagger_e(\mathbf{r})\hat{\Psi}_g(\mathbf{r})\rho_{00}\hat{\Psi}^\dagger_g(\mathbf{r}''')\hat{\Psi}_e(\mathbf{r}''')
	                   \hat{\Psi}^\dagger_e(\mathbf{r}^{'})\hat{\Psi}_g(\mathbf{r}^{'})\hat{\Psi}^\dagger_g(\mathbf{r}'')\hat{\Psi}_e(\mathbf{r}'')	                   
	                  \nonumber \\
	                  & & -\int d^3 rd^3r^{'}d^3r^{''}d^3r^{'''} \, \frac{L^{*}(\mathbf{r}''-\mathbf{r}')\Omega^*(\mathbf{r})
		             \Omega(\mathbf{r}^{'''})}{\Delta^2} 
	                  \hat{\Psi}^\dagger_e(\mathbf{r})\hat{\Psi}_g(\mathbf{r})\hat{\Psi}^\dagger_g(\mathbf{r}''')\hat{\Psi}_e(\mathbf{r}''')
	                     \rho_{11}
	                   \hat{\Psi}^\dagger_e(\mathbf{r}^{'})\hat{\Psi}_g(\mathbf{r}^{'})\hat{\Psi}^\dagger_g(\mathbf{r}'')\hat{\Psi}_e(\mathbf{r}'')	                   
	                  \nonumber \\
	                   & & +i\int d^3 rd^3r^{'} \, L(\mathbf{r}-\mathbf{r}') 
	                  \hat{\Psi}^\dagger_e(\mathbf{r})\hat{\Psi}_g(\mathbf{r})\hat{\Psi}^\dagger_g(\mathbf{r}')\hat{\Psi}_e(\mathbf{r}')\rho_{11}
	                   +H.c..
     \end{eqnarray}
\end{widetext} 	
Nextt, we apply a similar adiabatic elimination process eliminate the excited state matrix $\rho_{11}$. To do that, first we normal order the ground state operators in above equation, and then by setting $\dot{\rho}_{11}\approx 0$, obtains
\begin{widetext}
\begin{eqnarray}
\label{rho11-solution}
 	\rho_{11}&=& i\int d^3 r d^3 r^{'} \,\frac{\Omega^{*}(\mathbf{r})\Omega(\mathbf{r}')}{2\Delta\Gamma}
	                   \hat{\Psi}^\dagger_e(\mathbf{r})\hat{\Psi}_g(\mathbf{r}) \rho_{00}  \hat{\Psi}^\dagger_g(\mathbf{r}')\hat{\Psi}_e(\mathbf{r}')
	                   -i\int d^3 r d^3 r^{'} \,\frac{\Omega^{*}(\mathbf{r})\Omega(\mathbf{r}')}{2\Delta\Gamma}
	                   \hat{\Psi}^\dagger_e(\mathbf{r})\hat{\Psi}^\dagger_g(\mathbf{r}')\hat{\Psi}_g(\mathbf{r})\hat{\Psi}_e(\mathbf{r}')\rho_{11} 
	                  \nonumber \\
	                  & & -i\int d^3r \frac{|\Omega(\mathbf{r})|^2}{2\Delta\Gamma}\hat{\Psi}^\dagger_e(\mathbf{r})\hat{\Psi}_e(\mathbf{r})\rho_{11}-
                          \int d^3 rd^3r^{'} \, \frac{L(\mathbf{r}-\mathbf{r}')}{2\Gamma} 
	                  \hat{\Psi}^\dagger_e(\mathbf{r})\hat{\Psi}^\dagger_g(\mathbf{r}')\hat{\Psi}_g(\mathbf{r})\hat{\Psi}_e(\mathbf{r}')\rho_{11} 
	                  \nonumber \\
	                    & & +\int d^3 rd^3r^{'}d^3r^{''} \, \frac{L^{*}(\mathbf{r}''-\mathbf{r}')\Omega^{*}(\mathbf{r})
		             \Omega(\mathbf{r}^{'})}{2\Delta^2\Gamma} 
	                  \hat{\Psi}^\dagger_e(\mathbf{r})\hat{\Psi}_g(\mathbf{r})\rho_{00}\hat{\Psi}^\dagger_g(\mathbf{r}')\hat{\Psi}_g(\mathbf{r}')
	                   \hat{\Psi}^\dagger_g(\mathbf{r}^{''})\hat{\Psi}_e(\mathbf{r}^{''})                   
	                  \nonumber \\
	                  & & -\int d^3 rd^3r^{'}d^3r^{''}d^3r^{'''} \, \frac{L^{*}(\mathbf{r}''-\mathbf{r}')\Omega^*(\mathbf{r})
		             \Omega(\mathbf{r}^{'''})}{2\Delta^2\Gamma} 
	                  \hat{\Psi}^\dagger_e(\mathbf{r})\hat{\Psi}_g(\mathbf{r})\hat{\Psi}^\dagger_g(\mathbf{r}''')\hat{\Psi}_e(\mathbf{r}''')
	                     \rho_{11}
	                   \hat{\Psi}^\dagger_e(\mathbf{r}^{'})\hat{\Psi}_g(\mathbf{r}^{'})\hat{\Psi}^\dagger_g(\mathbf{r}'')\hat{\Psi}_e(\mathbf{r}'') 
	                   \nonumber \\	                   
	            	               & & +H.c.,
	 \end{eqnarray}
\end{widetext} 	
Again, we solve this equation perturbatively up to order $\frac{1}{\Delta^2}$, and arrive at
\begin{widetext}
\begin{eqnarray}
 	\rho_{11}&=& \int d^3 rd^3r^{'}d^3r^{''} \, \frac{L^{*}(\mathbf{r}''-\mathbf{r}')\Omega^{*}(\mathbf{r})
		             \Omega(\mathbf{r}^{'})}{2\Delta^2\Gamma} 
	                  \hat{\Psi}^\dagger_e(\mathbf{r})\hat{\Psi}_g(\mathbf{r})\rho_{00}\hat{\Psi}^\dagger_g(\mathbf{r}')\hat{\Psi}_g(\mathbf{r}')
	                   \hat{\Psi}^\dagger_g(\mathbf{r}^{''})\hat{\Psi}_e(\mathbf{r}^{''})                   
	                  \nonumber \\
	                  & & -\int d^3 rd^3r^{'}d^3r^{''}d^3r^{'''}d^3r'''' \, \frac{L(\mathbf{r}-\mathbf{r}')L^{*}(\mathbf{r}''''-
	                   \mathbf{r}''')\Omega^*(\mathbf{r}'')
		             \Omega(\mathbf{r}^{'''})}{2\Delta^2\Gamma^2} 
	                    \hat{\Psi}^\dagger_e(\mathbf{r})\hat{\Psi}^\dagger_g(\mathbf{r}')\hat{\Psi}_g(\mathbf{r})\hat{\Psi}_e(\mathbf{r}')
			     \hat{\Psi}^\dagger_e(\mathbf{r}'')\hat{\Psi}^\dagger_g(\mathbf{r}'')\rho_{00}\times   \nonumber \\
			     & &  \hat{\Psi}^\dagger_g(\mathbf{r}^{'''})\hat{\Psi}_g(\mathbf{r}^{'''})\hat{\Psi}^\dagger_g(\mathbf{r}'''')\hat{\Psi}_e(\mathbf{r}'''') 
	                   -\int d^3 rd^3r^{'}d^3r^{''}d^3r^{'''}d^3r'''' \, \frac{L(\mathbf{r}-\mathbf{r}')L^{*}(\mathbf{r}''''-
	                    \mathbf{r}''')\Omega^*(\mathbf{r}''')
		             \Omega(\mathbf{r}^{''})}{2\Delta^2\Gamma^2}\times      \nonumber \\	
	                    & &  \hat{\Psi}^\dagger_e(\mathbf{r})\hat{\Psi}^\dagger_g(\mathbf{r}')\hat{\Psi}_g(\mathbf{r})\hat{\Psi}_e(\mathbf{r}')
		        	\hat{\Psi}^\dagger_e(\mathbf{r}'''')\hat{\Psi}^\dagger_g(\mathbf{r}'''')
			      \hat{\Psi}^\dagger_g(\mathbf{r}^{'''})\hat{\Psi}_g(\mathbf{r}^{'''})\rho_{00}
			       \hat{\Psi}^\dagger_g(\mathbf{r}'')\hat{\Psi}_e(\mathbf{r}'') 
	             	        +H.c..
	 \end{eqnarray}
\end{widetext} 
This express for $\rho_{11}$ can be simplified by normal ordering. Finally, we insert the normal ordered $\rho_{01},\rho_{01},\rho_{11}$ back into Eq. (\ref{rho00}), and keep only terms up to $\frac{1}{\Delta^2}$. Meanwhile, since the number density of atoms in BEC is small, we drop any 3-body and above terms. With these two assumptions, we obtain the resulting master equation (\ref{mastereq}), 
  \begin{widetext}
  \begin{eqnarray}
  \label{final-master}
	\dot\rho_s(t) &=& -iH_s\rho_s(t)-
                \frac{|\Omega|^2}{\Delta^2}\Gamma \int d^3r \hat{\Psi}^\dagger_g(\mathbf{r})\hat{\Psi}_g(\mathbf{r})\rho_s(t)- \nonumber \\
                     & &  \int d^3r d^3r' \, \frac{\Omega(\mathbf{r})L(\mathbf{r}-
                    \mathbf{r}')\Omega^*(\mathbf{r}')}{\Delta^2}\hat{\Psi}^\dagger_g(\mathbf{r})\hat{\Psi}^\dagger_g(\mathbf{r})
                    \hat{\Psi}_g(\mathbf{r'})\hat{\Psi}_g(\mathbf{r}')\rho_s(t)+\nonumber \\  
                    & & \int d^3r d^3r' \frac{\Omega^*(\mathbf{r})L(\mathbf{r}-
                    \mathbf{r}')\Omega(\mathbf{r}')}{\Delta^2}
           \hat{\Psi}^\dagger_g(\mathbf{r})\hat{\Psi}_g(\mathbf{r})\rho_s(t)\hat{\Psi}^\dagger_g(\mathbf{r'})\hat{\Psi}_g(\mathbf{r}') +H.c.. \nonumber
 \end{eqnarray}
 \end{widetext}
where we have written $\rho_s \equiv \rho_{00}$.

\end{document}